\documentclass{aa}
\usepackage{txfonts}
\usepackage{natbib, twoopt}
\usepackage{float}
\usepackage{graphicx}
\usepackage[caption=false]{subfig}
\usepackage{color}
\usepackage{booktabs}

\begin{document}


	\title{IRIS observations of the Mg \textsc{ii} h \& k lines during a solar flare}
	
	\author{G.~S.~Kerr\inst{1} \and P.~J.~A. Sim{\~o}es\inst{1} \and J.~ Qiu\inst{2} \and L.~Fletcher\inst{1}}
	
	\institute{SUPA, School of Physics and Astronomy, University of Glasgow, G12 8QQ, Scotland, U.K. \\ \email{g.kerr.2@research.gla.ac.uk}
     			 \and Department of Physics, Montana State Univeristy, Bozeman Mt, USA
			 }
			 
	\date{Received / Accepted}
	
\keywords{Sun: flares - Sun: chromosphere - Sun: UV radiation}

	\abstract{The bulk of the radiative output of a solar flare is emitted from the chromosphere, which produces enhancements in the optical and UV continuum, and in many lines, both optically thick and thin. We have, until very recently, lacked observations of two of the strongest of these lines: the Mg \textsc{ii} h \& k resonance lines.  We present a detailed study of the response of these lines to a solar flare. The spatial and temporal behaviour of the integrated intensities, k/h line ratios, line of sight velocities, line widths and line asymmetries were investigated during an M class flare (SOL2014-02-13T01:40). Very intense, spatially localised energy input at the outer edge of the ribbon is observed, resulting in redshifts equivalent to velocities of $\sim$15-26km/s, line broadenings, and a blue asymmetry in the most intense sources. The characteristic central reversal feature that is ubiquitous in quiet Sun observations is absent in flaring profiles, indicating that the source function increases with height during the flare. Despite the absence of the central reversal feature, the k/h line ratio indicates that the lines remain optically thick during the flare. Subordinate lines in the Mg \textsc{ii} passband are observed to be in emission in flaring sources, brightening and cooling with similar timescales to the resonance lines. This work represents a first analysis of potential diagnostic information of the flaring atmosphere using these lines, and provides observations to which synthetic spectra from advanced radiative transfer codes can be compared.}
		
		\maketitle

	\section{Introduction}\label{sec:intro}
	 Solar flares are dramatic energy release events in the Sun's atmosphere, during which an enormous amount of free magnetic energy is liberated from the reconfiguration of the coronal magnetic field (commonly thought to be via magnetic reconnection). This liberated energy causes particle acceleration, heating and mass motions in the solar atmosphere, which results in a broadband enhancement to the solar radiative output. The bulk of this radiative output originates from the relatively dense chromosphere (e.g \citealt{2010MmSAI..81..616F, 2014ApJ...793...70M} and references therein), so observations of the flaring chromosphere contain diagnostics important to the understanding of energy and radiation transport during flares. 

	The chromosphere emits strongly in several ultraviolet (UV) and optical lines, but routine spectral observations have been largely confined to optical lines accessible from the ground (e.g. the H$\alpha$ line, and the resonance lines of Ca \textsc{ii}).  The strong UV Mg \textsc{ii} resonance lines have been rarely observed and only once, to our knowledge, during a solar flare \citep{1984SoPh...90...63L}. With the launch of the Interface Region Imaging Spectrograph (IRIS; \citealt{2014SoPh..289.2733D}) we are in a position to routinely observe the diagnostic rich Mg \textsc{ii} lines during flares, with excellent spatial and spectral resolution. 
	
\begin{figure}
			\vbox{
			\hbox{
			\subfloat[Quiet Sun MgII Spectrum]{\label{fig:av_mgii}
				\resizebox{\hsize}{!}{\includegraphics{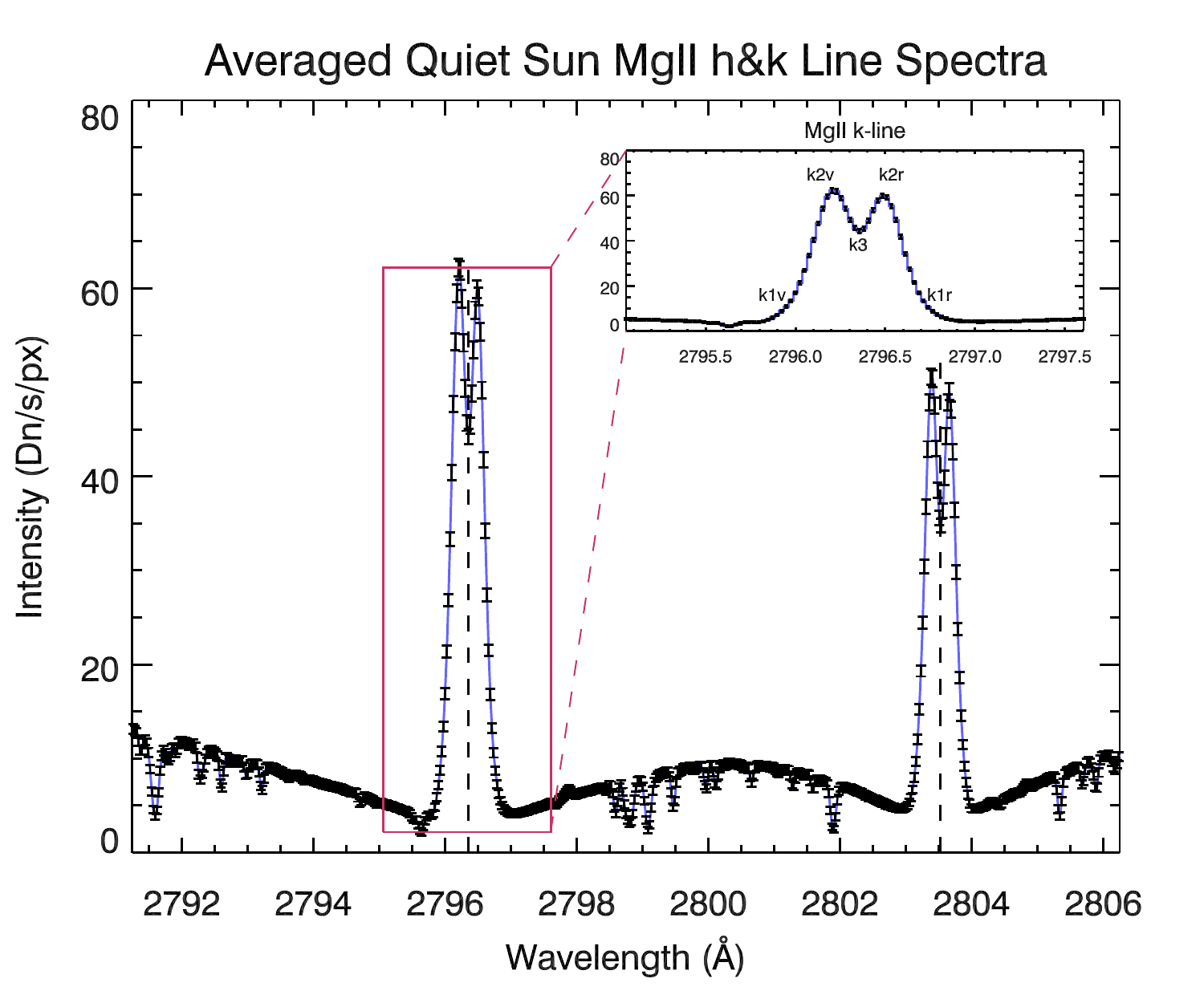}}}
				}
			\hbox{
			\subfloat[Example SJI image]{\label{fig:SJI_slitpos}
				\resizebox{\hsize}{!}{\includegraphics{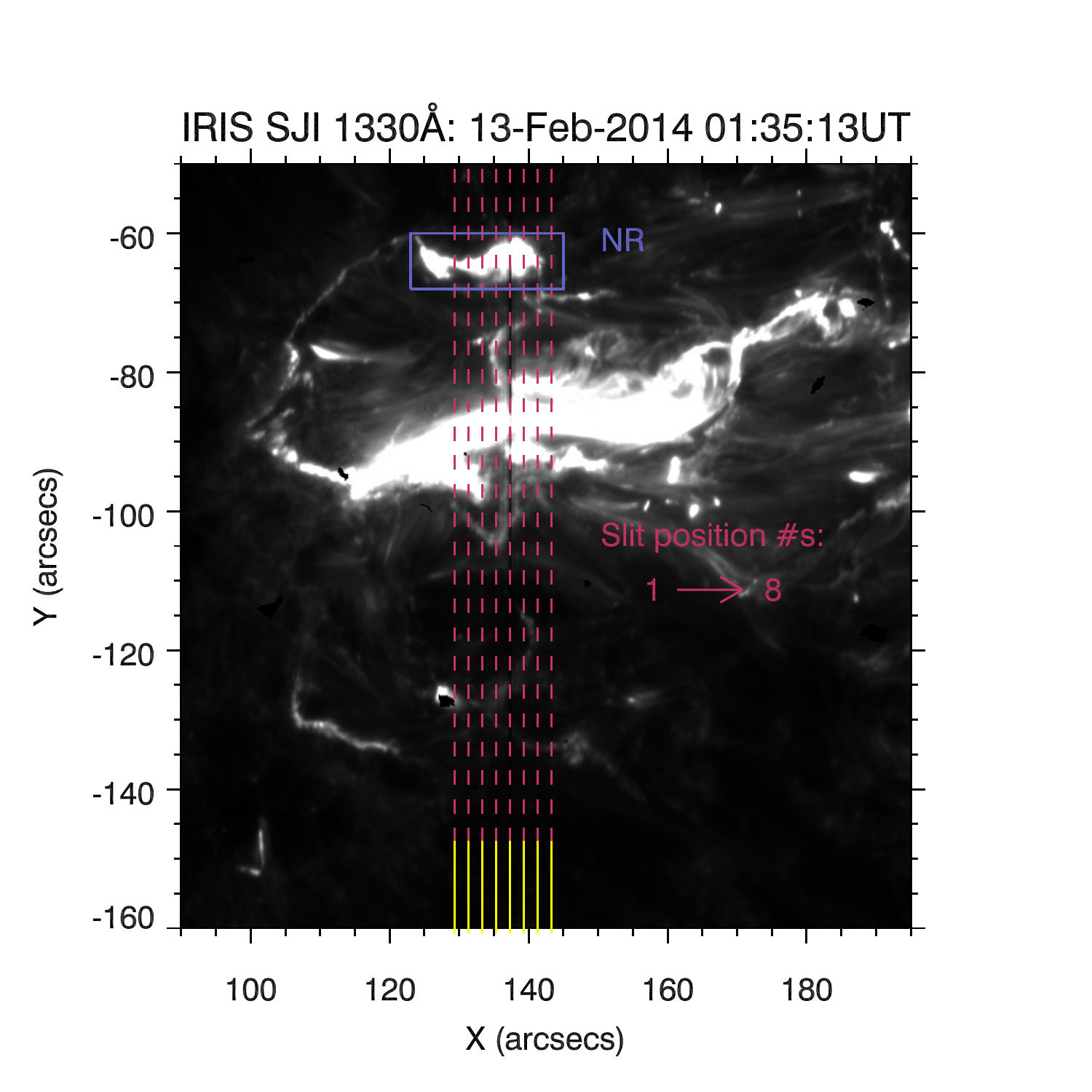}}}
				}
				}
			\caption{\textsl{(a) Quiet Sun Mg \textsc{ii} h\&k spectra, averaged over $\approx$ 80000 pixels. The inset shows the components of the k-line (h-line labelled using the same convention). Dotted lines show the rest wavelengths: 2796.35\AA\ (k-line) and 2803.53\AA\ (h-line) (b) 1330\AA\ SJI from 13:35~UT. The blue box surrounds the northern ribbon (NR), the red dotted lines are the slit positions, and the yellow regions are the pixels that were used to create the averaged quiet Sun spectra.}}
			\label{fig:ex_sji_spectra}
		\end{figure}

	The Mg \textsc{ii} h \& k resonance lines are the $3s-3p$ transitions to the ground state of singly ionised magnesium, from upper levels that are close in energy ($3s~^2S_{1/2}-3p~^2P_{1/2}$, and $3s~^2S_{1/2}-3p~^2P_{3/2}$ respectively). They are optically thick lines that show a complex line profile, with three components identified. Figure~\ref{fig:ex_sji_spectra}(a) shows these components and indicates the naming convention: k3 (central reversal), k2v \& k2r (the violet and red emission peaks), and k1v \& k1r (the violet and red minima), with the h line similarly named. When discussing common properties or nomenclature between the resonance lines we will refer only to the k line profiles, unless otherwise noted. These lines have vacuum wavelengths in the near ultraviolet of 2796.34\AA\ (k-line) and 2803.52\AA\ (h-line).
	
	Several authors have previously commented on the excitation mechanisms of the resonance lines, (e.g \citealt{1982ApJS...49..293L, 1977ApJ...212L.147F, 2013ApJ...772...89L, 1983A&A...125..241L}). Electron impact excitation from the ground state has been found to be the dominant mechanism to populate the upper levels of the h \& k lines and subordinate lines. Higher energy levels are populated from the ground state and cascade through the lower levels of Mg~\textsc{ii}, populating the $3p$ levels which then radiatively de-excite back to the ground state. Photoionisation and direct radiative recombinations are negligible effects in comparison \citep{1982ApJS...49..293L}. 
			
	In addition to the resonance lines, a set of subordinate triplet lines is present in the vicinity of the resonance lines. These are the $3p^2P-3d^2D$ transitions of Mg~\textsc{ii}, with wavelengths: 2791.60\AA, 2798.75\AA\ and 2798.82\AA. Generally these lines are in absorption in the non-flaring atmosphere, though they have been observed to go into emission during a solar flare \citep{1984SoPh...90...63L} and above the limb \citep{1977ApJ...212L.147F}. Recently \cite{2015ApJ...806...14P} have modelled these lines, noting that under certain circumstances these lines can be in emission on the non-flaring disk (c.f Section~\ref{sec:discuss} for further comment).

	While observations of the Mg~\textsc{ii} lines have been rare in comparison to other passbands, several spectra have been obtained in the past. Rocket experiments by \cite{1973A&A....22...61L} and \cite{1976ApJ...205..599K} provided the first high resolution spectra. These were followed by space-based observations using the NRL spectrograph on board Skylab, \citep{1977ApJS...35..471D, 1977ApJ...212L.147F}, and the \textsl{Orbiting Solar Observatory} 8 (OSO-8), which provided simultaneous observations of Mg \textsc{ii} h \& k, Ca \textsc{ii} H \& K, and Ly$\alpha$ \& $\beta$. Several detailed studies of various solar features were performed by \cite{1981A&A...103..160L}, \cite{1981SoPh...69..289K}, \cite{1982ApJS...49..293L}, \& \cite{1984SoPh...90...63L} amongst others using the OSO-8 data.

	These authors demonstrated variation in the profiles over different solar features (quiet Sun, plage, network, sunspots) and over the disk towards limb. The line intensity was observed to be greater over plage and in active regions with a smaller central reversal. Asymmetries between the k2v and k2r features were present in many of their spectra, which were interpreted as flows in the atmosphere. Profiles from 6$''$ above the limb contained no central reversal, and above 12$''$ the k:h intensity ratio was $\sim$2, indicating that the lines had become optically thin \citep{1977ApJS...35..471D, 1977ApJ...212L.147F}. Sunspots lacked a central reversal but the k/h ratio still implied an optically thick line \citep{1981SoPh...69..289K}. 
	
	Only one flare spectrum has been reported in the literature. Using the OSO-8 LPSP instrument \citep{1984SoPh...90...63L} measured pre-flare and flare profiles of the Mg \textsc{ii} h \& k, Ca \textsc{ii} H \& K, and Ly$\alpha$ \& $\beta$ lines, and inferred the downward propagation of energy during the flare. Flaring profiles were enhanced and broadened but it is unclear if the central reversal disappears entirely due to poor sampling. The 3p-3d subordinate lines of Mg \textsc{ii} appear in emission during the flare, with a factor $\sim 3$ intensity increase. The ratio of k/h intensity is stable during the flare at a value of 1.1 (whereas the Ca \textsc{ii} K/H ratio increases from 1.0 to 1.2, with the suggestion that the Ca \textsc{ii} opacity increases due to the flare.)

	Modelling of Mg~\textsc{ii} in the non-flaring chromosphere has been carried out by  \cite{1968SoPh....3..181A}, \cite{1974ApJ...192..769M}, \cite{1976ApJ...205..874A}, \cite{1997SoPh..172..109U}, \cite{2013ApJ...772...89L,2013ApJ...772...90L} and \cite{2013ApJ...778..143P}. We summarise the basic formation process here but those authors should be consulted for an in-depth description.
		
	The h \& k lines form at multiple layers in the atmosphere, sampling regions from the temperature minimum to the upper chromosphere. Since they are optically thick we observe each frequency at its $\tau = 1$ height, and photons are able to escape. The k1 minima are formed near the temperature minimum region, with a source function that is still partially coupled to the Planck function. The k2 emission peaks form in the middle chromosphere, where their source function has decoupled from the Planck function but which has a maximum around the $\tau = 1$ height for those frequencies. Finally, the k3 emission cores form in the upper chromosphere (\citealt{2013ApJ...772...90L} finds the height to be less than 200~km below the transition region), where the source function has strongly departed from the Planck function, and which decreases with height resulting in the central reversal. The k line has a higher opacity than the h line by a factor of two and so forms slightly higher in the atmosphere (a few tens of km). This higher opacity also results in a stronger coupling to the gas temperature and a stronger source function (hence stronger emergent intensity) than the h line. Velocity and temperature gradients between the formation heights of the k2r and k2v components can result in asymmetric profiles and a variable separation of the emission peaks. 

	\cite{2013ApJ...772...89L, 2013ApJ...772...90L} \& \cite{2013ApJ...778..143P} modelled the Mg \textsc{ii} lines using a snapshot of the radiation magnetohydrodynamic (RMHD) code, Bifrost \citep{2011A&A...531A.154G} as an input atmosphere to a modified version of the radiative transfer code RH \citep{2001ApJ...557..389U}. This was an effort to develop diagnostics of the quiet chromosphere in preparation for the analysis of IRIS data. Their results were very interesting with several diagnostics available to discern atmospheric properties (the temperature and velocity structure).  However, it is not known if these diagnostics are applicable to active regions or flaring conditions. 
	
	With IRIS we can improve dramatically on the spatial and spectral resolution that was available to \cite{1984SoPh...90...63L}. We analysed Mg \textsc{ii} observations of a solar flare as the first step in improving our understanding of how these lines behave in the flaring chromosphere, and as a prelude to modelling these lines in a flaring atmosphere, in order to ascertain if they are as useful a diagnostic tool in flares as they have been shown to be in the quiet Sun. 
	
\begin{figure}
			\vbox{
			\hbox{
			\subfloat{\label{fig:rhessi_lc}
				\resizebox{\hsize}{!}{\includegraphics	{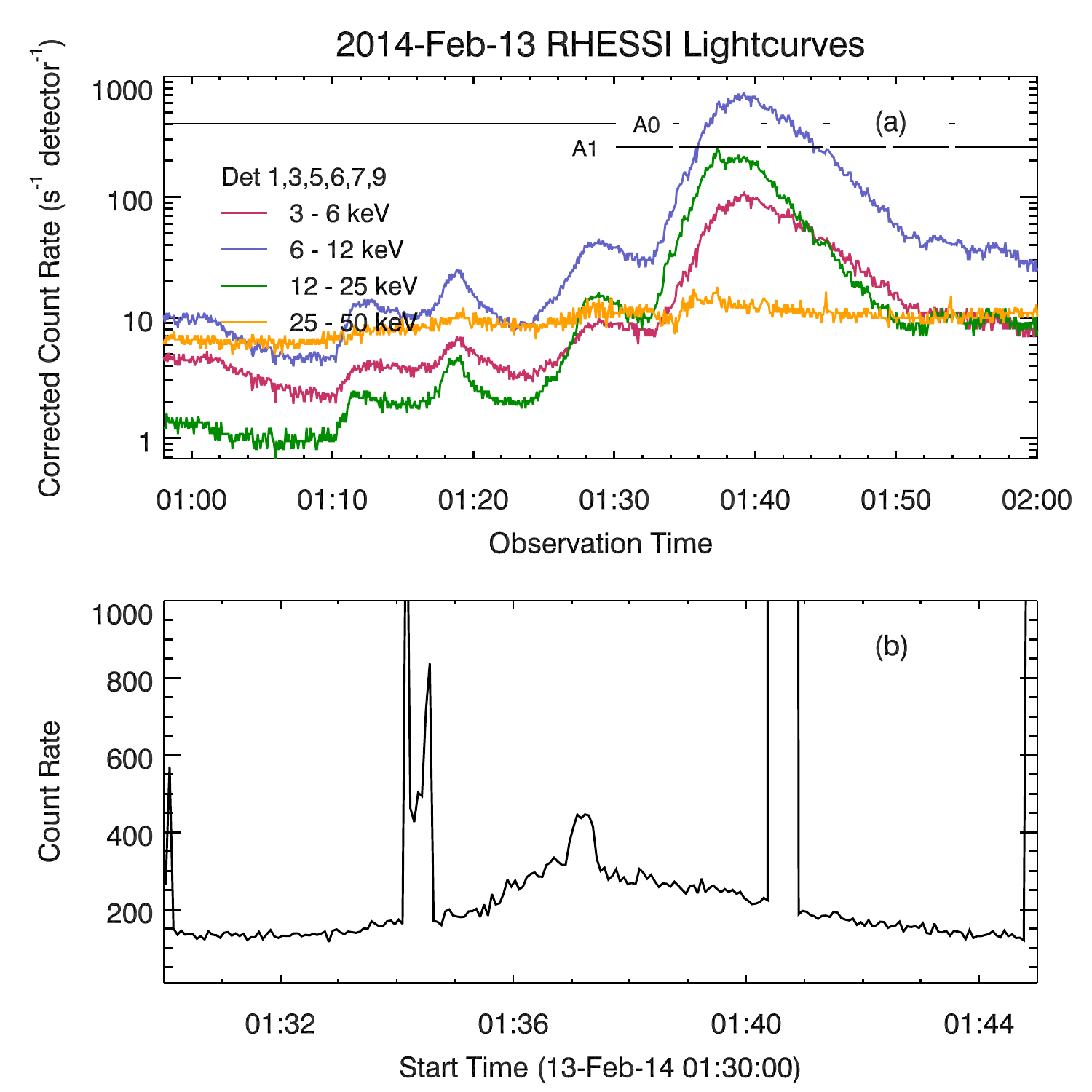}}}
			}
			\hbox{
			\subfloat{\label{fig:rhessi_spec}
				\resizebox{\hsize}{!}{\includegraphics{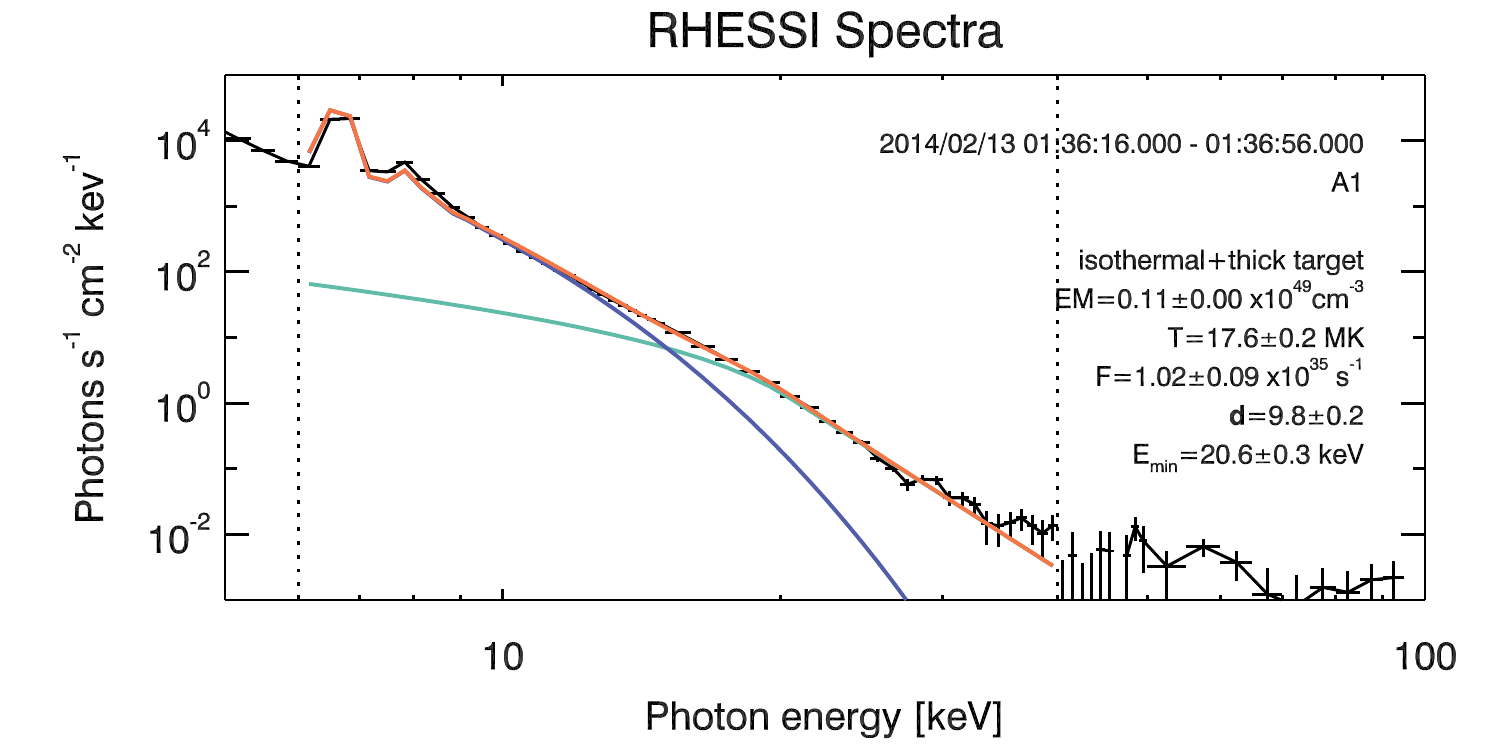}}}
			}
			}
			\caption{\textsl{a) RHESSI count rates (corrected for the attenuator changes, shown), for the 2014-Feb-13 M1.8 flare, also showing several pre-flare peaks. Dotted lines indicate the zoomed-in portion in the lower panel (b) The 20--50~keV count rate averaged over the detectors (note this has not been corrected for attenuator changes), illustrating the small non-thermal peak present at $\sim$ 01:35:30--01:38:00~UT. (c) RHESSI hard X-ray spectrum at 01:36:16~UT, fitted with an isothermal (green) plus thick-target (blue) model. Horizontal bars show the energy binning and vertical bars the uncertainty of the data. The vertical dashed lines indicate the range of the data used in the model fitting.}}
			\label{fig:hsi_figs}
		\end{figure}
	\section{Observations}\label{sec:obs}
	NOAA active region 11974 produced an M1.8 class solar flare (SOL2014-02-13T01:40) on 2014 Feb 13 that began $\approx$ 01:32UT, peaking at $\approx$ 01:38UT, accompanied by a failed filament eruption. This event, located at $\sim$[140, -90]$''$, was well observed by IRIS and by the Reuven Ramaty High-Energy Solar Spectroscopic Imager (RHESSI; \citealt{2002SoPh..210....3L}). We focus on the Mg \textsc{ii} h \& k spectral line data from IRIS, but present X-ray data from RHESSI as a guide for future modelling. We present some context images from the Atmospheric Imaging Assembly (AIA; \citealt{2012SoPh..275...17L}) onboard the Solar Dynamics Observatory (SDO; \citealt{sdo_paper}).

		\subsection{IRIS}\label{sec:IRISobs}
		IRIS is a NASA small explorer spacecraft observing the solar atmosphere with remarkable spatial and spectral resolution in the near- and far-ultraviolet (NUV \& FUV). Images are provided via the slit-jaw imager (SJI) and spectral data by the spectrograph (SG) that observes several channels through a slit of 0.33$''$ width. 
				
		For the event that we report on here IRIS was observing with an 8-step raster program. SJI images with a cadence of $\approx 11$s, a pixel scale of 0.167$''$~pixel$^{-1}$ and a field of view 119$''$ x 119$''$ were made in the 1330\AA\ and 1400\AA\ passbands. Spectra were observed in an 8-step sequence, stepping 2$''$ between exposures (4 slit positions simultaneous with 1330\AA\ images and 4 simultaneous with 1400\AA\ images), with a repeat cadence of 43s. The Mg~\textsc{ii} spectra had a pixel scale in the y-direction of 0.167$''$~pixel$^{-1}$ and spectral resolution 25.46 m\AA\ pix$^{-1}$.
			
		We used the level 2 data product which has already been corrected for cosmic ray spikes, dark current, flat-fielding, geometry  and (as of the April 2014 pipeline) wavelength calibration. Wavelength calibration does not account for thermal variations during orbit or variations in orbital velocity, and so these were further corrected for using the \texttt{iris\_orbitvars\_corr\_l2.pro} routine, part of the IRIS \texttt{SolarSoftware} (SSW; \citealt{1998SoPh..182..497F}) tree. Dividing by exposure time provided intensity in DN~s$^{-1}$~px$^{-1}$. 
	
		The IRIS pointing was offset from the AIA and RHESSI data by several arcseconds. To correct for this the AIA 1600\AA\ maps, which share many of the same features as the IRIS 1330\AA\ \& 1400\AA\ passband were re-binned to the size of the 1330\AA\ maps, and the two data sets cross-correlated. The IRIS images were shifted to match the AIA images. 
		
		Figure~\ref{fig:ex_sji_spectra}(b) shows an example 1330\AA\ SJI image from near the peak of the flare. The eight dashed red lines indicate the slit positions and solid yellow lines indicate the pixels used to calculate an average quiet-Sun spectrum. The northern ribbon is clearly seen in this image, whereas the location of the southern ribbon is less clear, being obscured by the filament. An averaged quiet Sun profile of the Mg \textsc{ii} h \& k lines is shown in Figure~\ref{fig:ex_sji_spectra}(a), where spectra have been averaged along each of the 8 slit positions and over time (from $\approx$ 01:00UT to $\approx$ 02:30UT), for the patch of quiet Sun indicated in Figure~\ref{fig:ex_sji_spectra}(b). Using the quartiles analysis described in Section~\ref{sec:quartiles}, the line centroid values for the h \& k lines were measured for each profile. The line centroid wavelengths of the quiet Sun pixels indicated in Figure~\ref{fig:ex_sji_spectra}(b) were averaged to give the resonance line rest wavelengths. These were $\lambda_\mathrm{k,rest} = 2796.3292$\AA\ and $\lambda_\mathrm{h,rest} = 2803.5181$\AA, with a standard deviation of $\pm$0.0032\AA. Vacuum rest wavelengths of 2796.35\AA\ and 2803.53\AA\ are quoted in the SSW routine \texttt{irisspectobs\_\_define.pro}, slightly larger than those measured. We used the quiet-Sun-derived rest wavelengths in our analysis. These are overlaid on Figure~\ref{fig:ex_sji_spectra}(a). The complicated profile of the Mg \textsc{ii} resonance lines can be seen in this quiet-Sun spectrum, with the central reversal (k3 \& h3 components) and the two emission peaks (k2 \& h2 components). 
	 \begin{figure}
		 	\centering
			\vbox{
			\hbox{
			\subfloat{\label{fig:aia_rhessi_sxr}}
				\includegraphics[width = 0.42\textwidth]{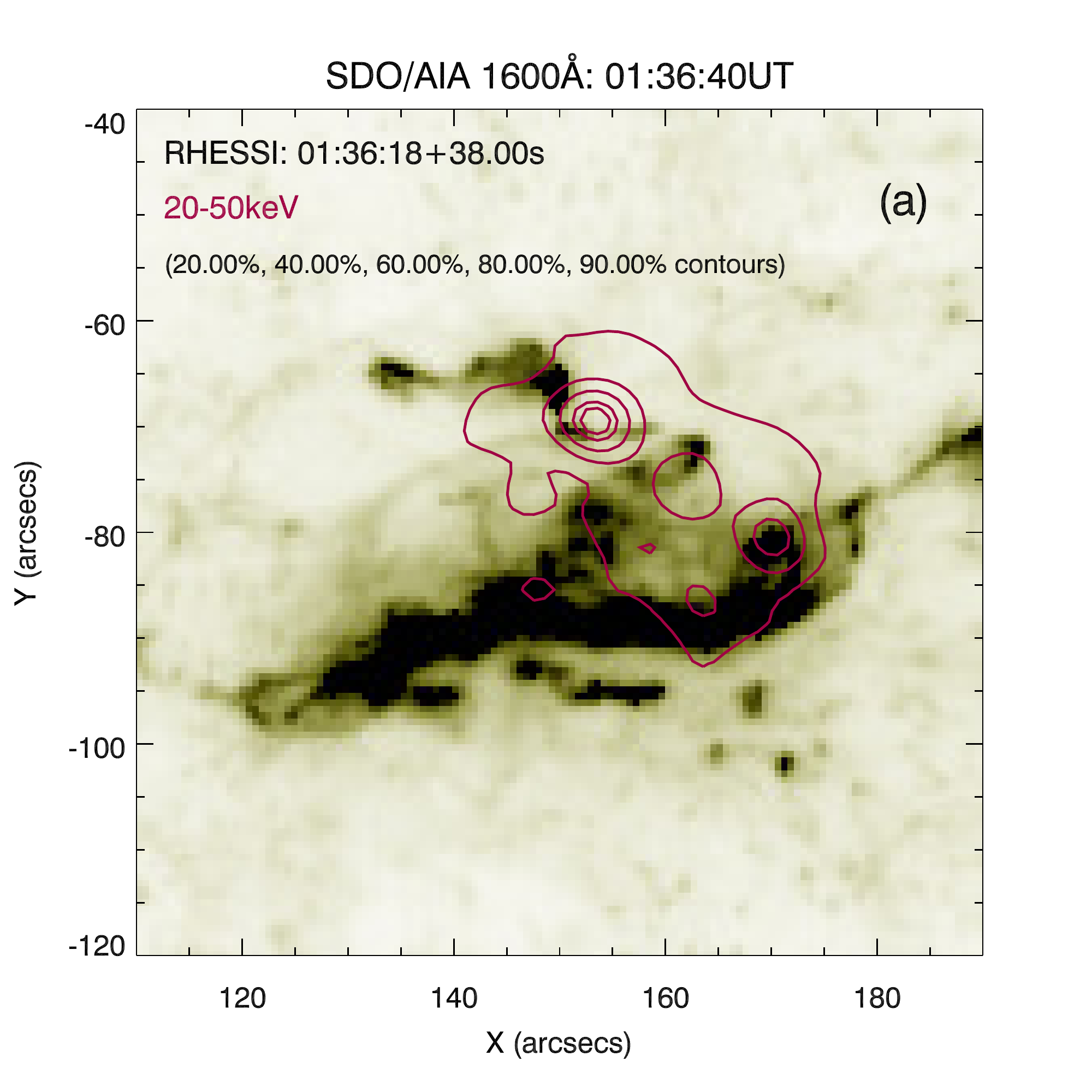}
			}
			\vspace{-0.55cm}
			\hbox{
			\subfloat{\label{fig:aia_rhessi_hxr}}
				\includegraphics[width = 0.42\textwidth]{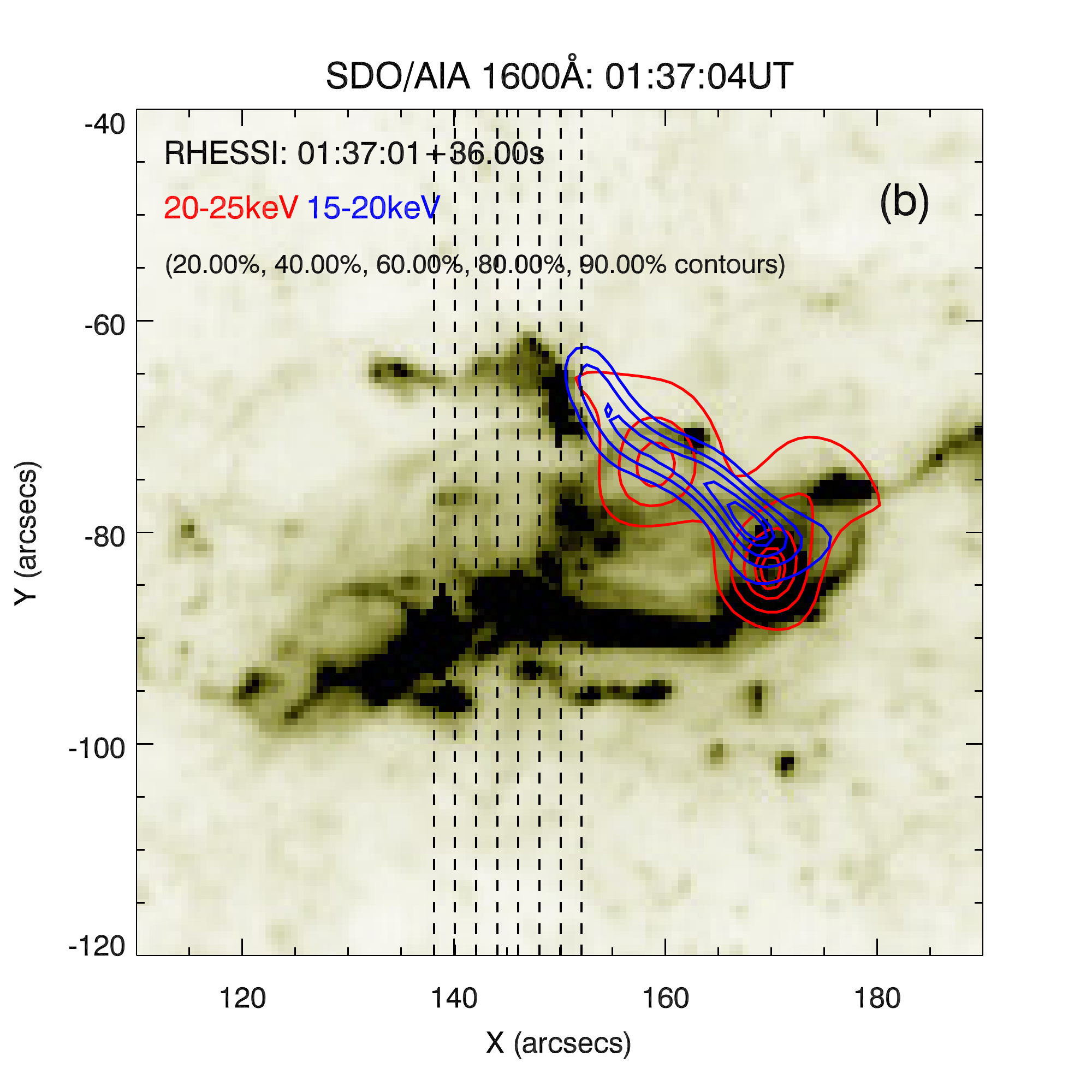}
			}
			\vspace{-0.55cm}
			\hbox{
			\subfloat{\label{fig:aia_rhessi_nth}}
				\includegraphics[width = 0.42\textwidth]{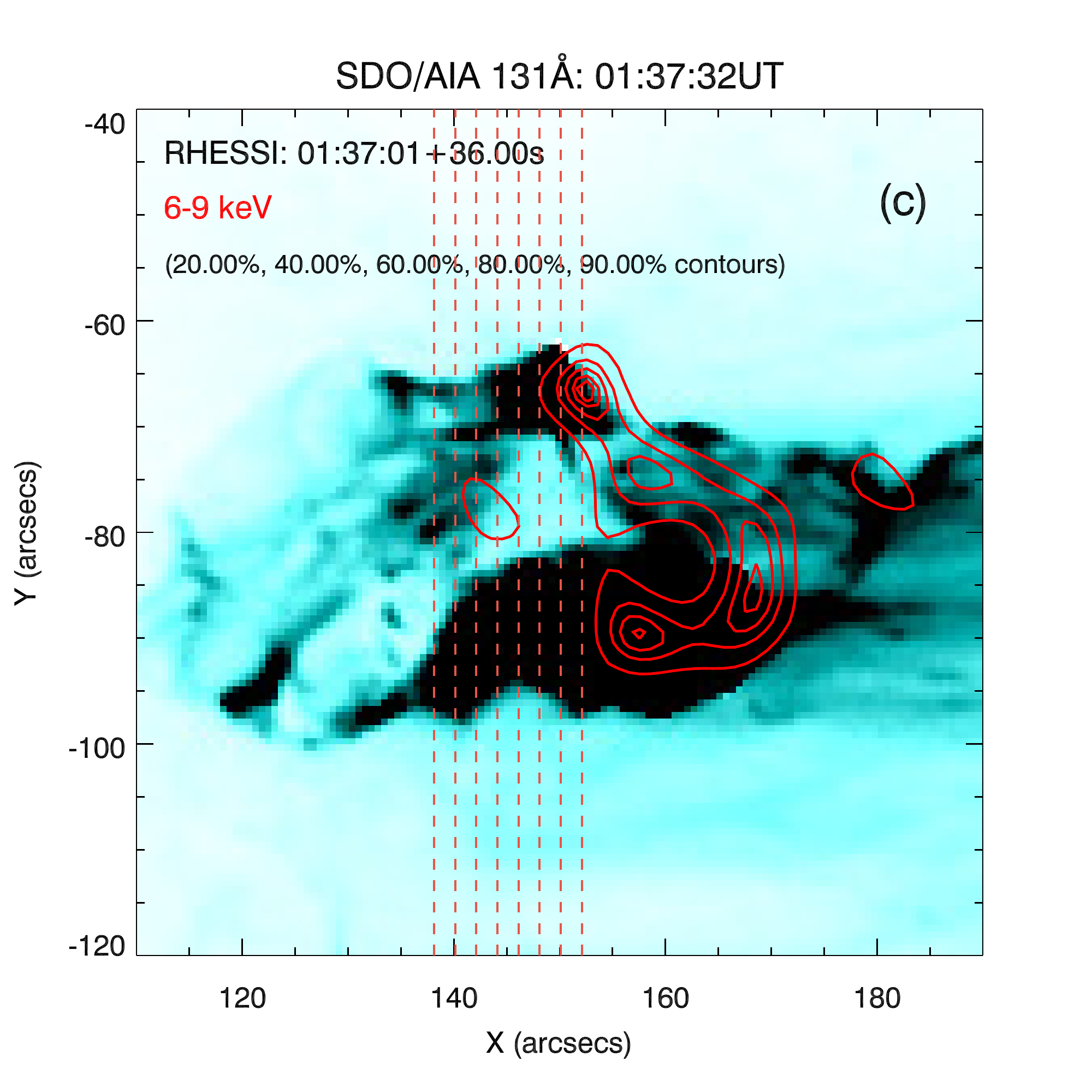}
			}
			}
			\caption{\textsl{These images indicate the spatial association of the RHESSI X-ray sources with the ribbon structure. (a) shows the only time at which a non-thermal source (20--50~keV) was near the northern ribbon structure. (b) shows that a short time later, the higher energy source are no longer located near the ribbon (this image uses 20--25~keV and 15-20~keV as the higher energy passbands contained significant noise at these times). (c) shows the 6--9~keV compact sources that are positioned near the northern ribbon. Dotted lines show the IRIS slit positions.}}
			\label{fig:aia_sequence}
		\end{figure}

		The error on intensity, $I$ measured in Data Numbers (DN) is a combination of the photon counting error and the readout noise/dark current uncertainty, which were added in quadrature. The readout noise is stated as $\sigma_{\mathrm{rn}} = 1.2$~DN by \cite{2014SoPh..289.2733D}. To convert DN to photons, this value was multiplied by the gain, the number of photons per DN, which is $g = 18$ for the NUV \citep{2014SoPh..289.2733D}. For the NUV channel, the fractional error on intensity measured in photons was calculated as 
	
		\begin{equation}
			f_\mathrm{err} = \frac{\sigma_{\mathrm{phot}}}{gI} = \frac{\sqrt{gI + (g \sigma_{\mathrm{rn}})^2}}{gI}
		\end{equation}

		Using $f_\mathrm{err}$, the error on $I$ measured in DN, and DN~s$^{-1}$ was calculated.
			
		\subsection{RHESSI}\label{sec:RHESSobs} 	
		RHESSI, launched in 2002, is designed to investigate the hard X-ray emission of solar flares through imaging \citep{2002SoPh..210...61H} and spectroscopy \citep{2002SoPh..210..165S}. RHESSI observed the rise phase and peak of the flare as well as most of the initial decay. Figure~\ref{fig:hsi_figs}(a) shows the RHESSI lightcurves for this event, corrected for changes in attenuator state (thin attenuattor (state ``A1") was on during most of the main phase of this event). It is clear that there is not a strong high-energy signal, though Figure~\ref{fig:hsi_figs}(b) shows that a weak peak is observable. Here we have summed the counts in the 20--50~keV channels which were not corrected for attenuator changes (the large spikes are due to changing attenuator state).  	
		
		We performed RHESSI spectral analysis for various time intervals over the peak of the event (time are indicated in Table~\ref{tab:rh_params}), using standard \texttt{OSPEX} software \citep{2002SoPh..210..165S}. The HXR spectrum was fitted with an isothermal plus thick-target model. Assuming the non-thermal electron distribution is a single power-law of the form 
		
		\begin{equation}
			F=\frac{(\delta-1)}{E_\mathrm{min}^{1-\delta}}\int_{E_\mathrm{min}}^\infty E^{-\delta} dE,
		\end{equation}
		
		the thick-target model parameters were found. These are the Flux, $F$, the total electron rate above the low energy cutoff $E_c$, and spectral index, $\delta$. N.b. $E_\mathrm{min}$ is the highest value of the low energy cutoff consistent with the spectrum, and the electron rate is correspondingly a lower limit. Table~\ref{tab:rh_params} shows these values at several time intervals, along with the total power in the electron distribution. The high values of $\delta$ indicate a very soft spectrum (weak non-thermal emission). Figure~\ref{fig:hsi_figs}(c) shows the fit for the time interval 01:36:16--01:35:56~UT.

		X-ray images were constructed in the 20--50~keV energy range, integrated over 38s time ranges around the flare peak by applying the imaging algorithm MEM NJIT \citep{2007SoPh..240..241S}. Of these, only the images from 01:36:18 to 01:36:56~UT showed sources near the flare ribbons. For the x-ray peak (01:37:01 to 01:37:37 UT), images were also constructed in 6--9~keV, 15--20~keV, \& 20--25~keV ranges. These images are shown as contours in Figure~\ref{fig:aia_sequence}.\\
		
\begin{table}	
	\centering
	\caption{RHESSI Fit Parameters}
	\begin{tabular}{lcccc} 
		\toprule
 			\multicolumn{4}{r}{Parameters} \\ 
 			\cmidrule(r){2-5} 
			Time & $\delta$ & $E_c$ & Flux  & P \\ 
			 & & (keV) & 10$^{35}$ elec. $s^{-1}$) & (erg $s^{-1}$) \\
			 \midrule  
			  01:34:52~UT+36$s$ & 9.3 & 19.6  & 0.46 & 1.62$\times 10^{27}$ \\ 
			  01:35:36~UT+36$s$ & 8.6 & 19.5  & 0.75 & 2.69$\times 10^{27}$ \\
			  01:36:16~UT+40$s$ & 9.8 &  20.6 & 1.02 & 3.77$\times 10^{27}$ \\
			  01:37:00~UT+44$s$ & 8.9 &  20.1 & 1.26 & 4.62$\times 10^{27}$ \\		  
  			\bottomrule
 	  \end{tabular} 
	  
	  \label{tab:rh_params}
\end{table}

	\subsection{Flare Overview}\label{sec:context}	
			
	 RHESSI lightcurves in Figure~\ref{fig:hsi_figs}(a) show that the rise phase of the flare begins at $\sim$01:32UT, and peaks at $\sim$01:37UT. There is pre-flare activity present in the lower energy channels, observed as small peaks at $\sim$01:10UT, $\sim$01:19UT, and $\sim$01:28UT. Pre-flare activity can also be seen in the AIA images, with hot loops visible in the coronal 131\AA\ and 193\AA\ channels that brighten concurrently with RHESSI pre-flare enhancements. 
	  
	   The flare ribbons, observed in UV images from AIA and IRIS SJI, begin to form at $\sim$01:32UT, with the northern ribbon (NR) clearly visible in all passbands. As the flare proceeds to the peak the ribbon expands northwards with a bright front which we call the `outer ribbon', leaving a less intense wake that decays gradually as it moves on that we refer to as the `inner ribbon'. The southern ribbon location is likely hidden behind the erupting filament. 

	Figure~\ref{fig:aia_sequence} shows the flaring region, with 1600\AA\ and 131\AA\ images from around the peak of the event. RHESSI X-ray sources are overlaid on the UV images. The non-thermal 20--50keV sources are only strongly visible and spatially associated with the NR source at $\sim$01:36:18~UT. Before this point, and afterwards, non-thermal emission is not present near flaring ribbons. Images from the peak x-ray time ($\sim$01:37:01~UT) show non-thermal 20--25~keV emission that is no longer spatially associated with the NR source (Figure~\ref{fig:aia_sequence}(b)). However, there are compact thermal 6--9keV footpoint sources that are spatially well associated with the UV ribbons (Figure~\ref{fig:aia_sequence}(c)). This appears to be a very thermal event with weak non-thermal signatures, that are only present for a short time. 

	   	 	 A filament is seen erupting in the AIA channels at $\sim$01:33:30UT, with the eruption and expansion of the filamentary material occurring over several minutes. The filament expands over the north half of the region eventually obscuring the northern flare ribbon (starting at $\sim$01:40UT and by $\sim$01:47UT the ribbon is completely obscured). This is not as obvious in the SJI in which the filament eruption results in a bright core of material.


	\section{Mg \textsc{ii} Analysis}\label{sec:MgIIanal}
	In the following sections we investigate the behaviour of the Mg \textsc{ii} resonance lines, focussing on the pixels in the NR as these are unambiguously flaring sources. Southern ribbon (SR) sources are easily confused with the filament, particularly during the eruption. 

	\begin{figure}
			\resizebox{\hsize}{!}{\includegraphics{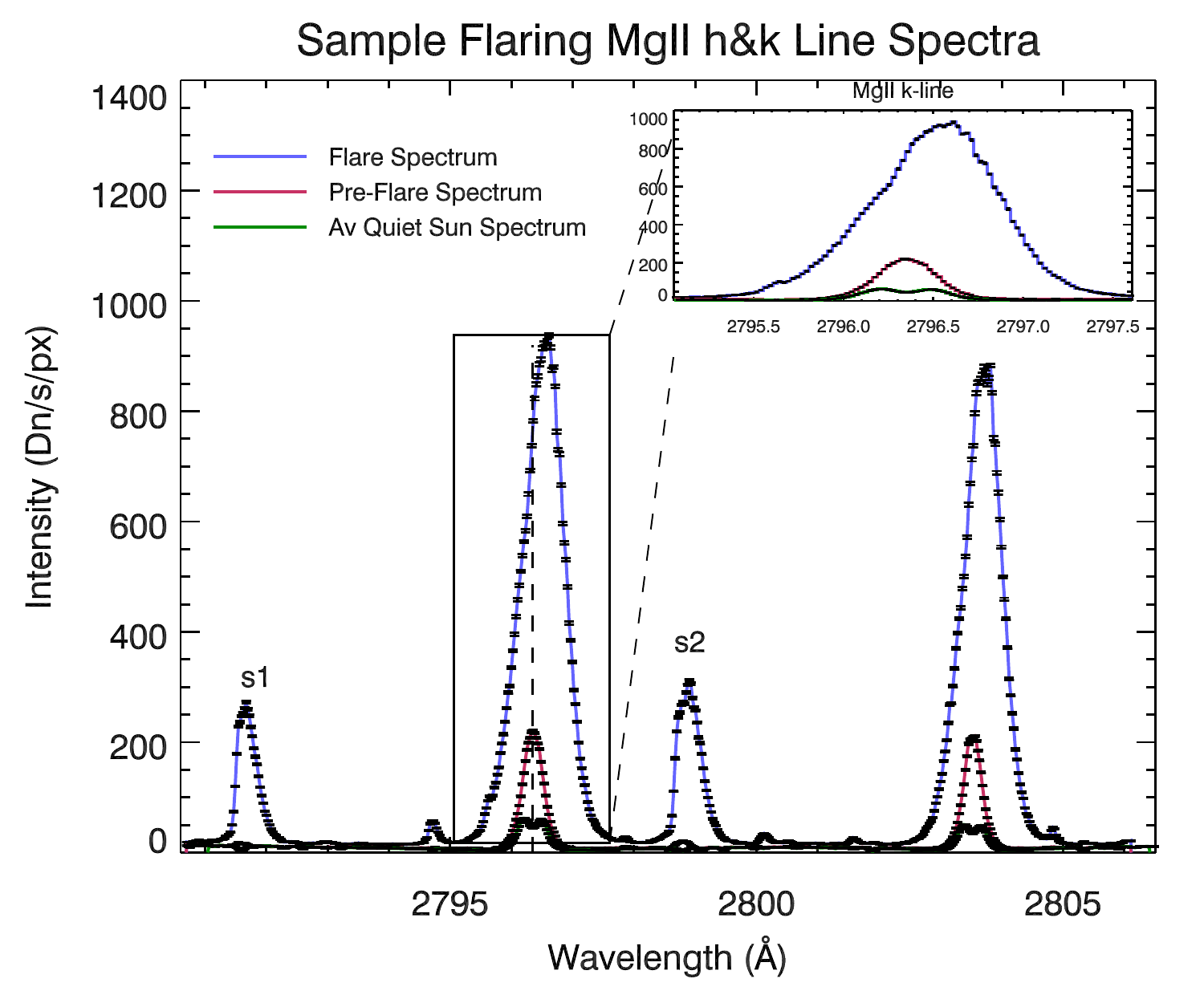}}	
			\caption{\textsl{A sample flare spectrum (blue), with the pre-flare (red) and averaged quiet Sun (green) spectra shown alongside. No central reversal is seen in the flare or pre-flare 								spectra.}}
			\label{fig:ex_flarespectra}
		\end{figure}
		
		\begin{figure}
			\centering
			\includegraphics[width=.5\textwidth]{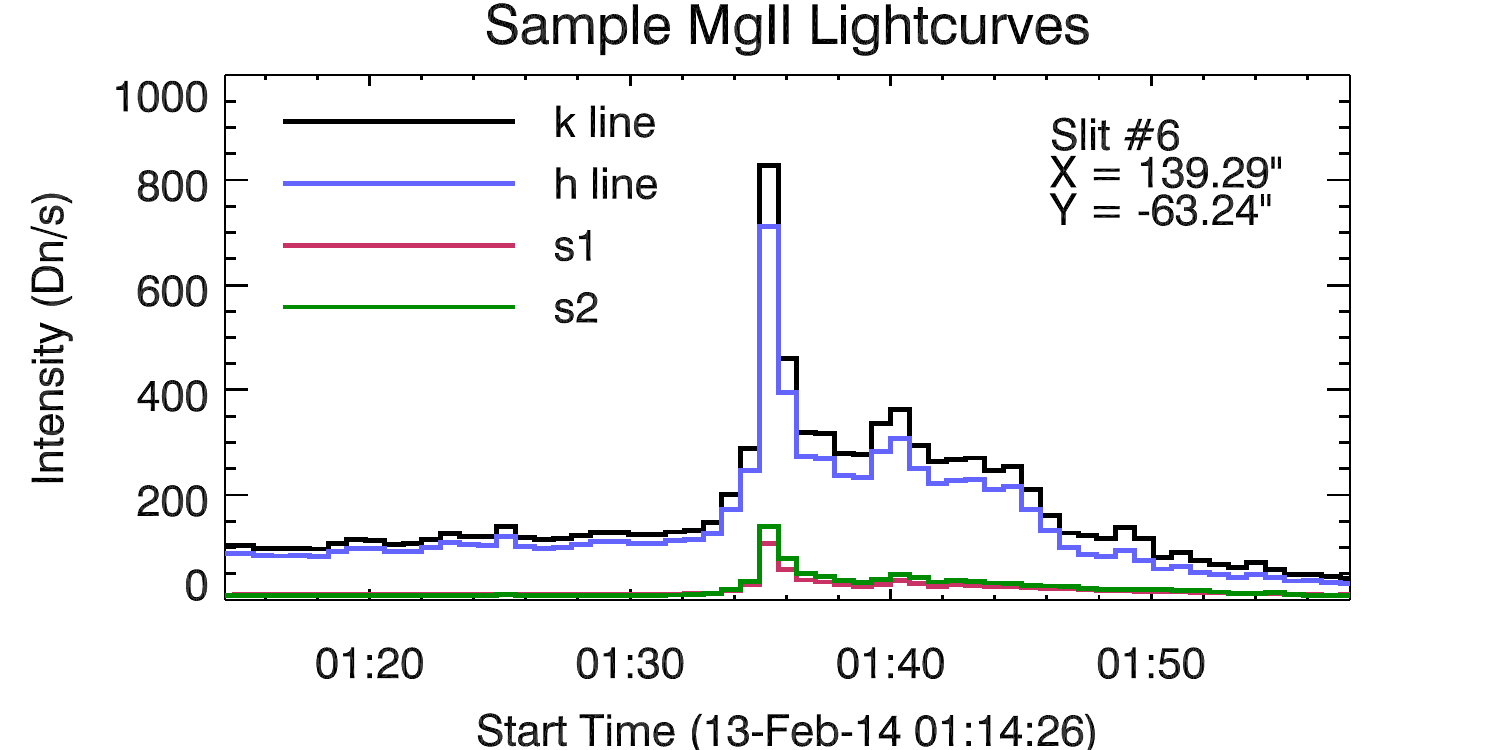}
			\caption{\textsl{A sample lightcurve showing the very impulsive peak.}}
			\label{fig:ex_lcurve}	
		\end{figure}

		\subsection{General behaviour}\label{sec:general}
		We first describe in general terms the behaviour during the flare. As discussed in Section~\ref{sec:context}, the northern ribbon is observed to spread northwards and then slightly to the west. Taking several cuts across the SG's 8 slit positions at the ribbon locations (Figure \ref{fig:ex_sji_spectra}(a)), the general behaviour of the Mg \textsc{ii} spectra during pre-flare times (from $\sim$30 mins before the flare), the time of initial energy deposition, and the time spent in the inner ribbon, was studied.
 
 		Before the flare, spectra in the region of the northern ribbon did not always have a central reversal. The line profiles appear symmetric (but not gaussian) and very rapidly.
		
		\begin{figure*}
			\includegraphics[width = 17cm]{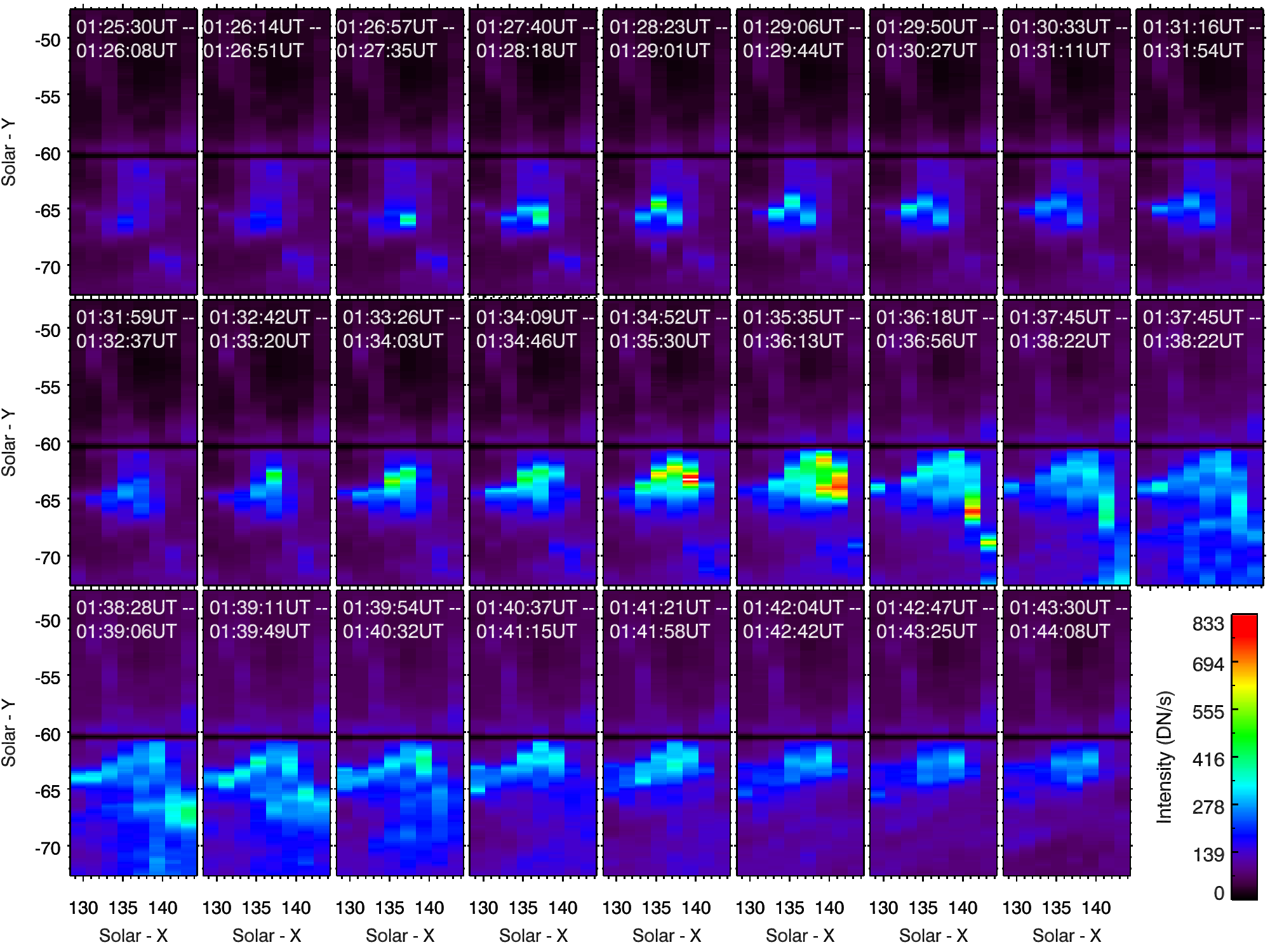}
				\caption{\textsl{A sequence of images showing the integrated intensity of the k line over time. Each panel is made up of 8 slit positions, with the start and end times of the scan indicated. The solid black line is a fiducial mark on the SG.}}
			\label{fig:int_intensity_sequence}
		\end{figure*}

		Shortly before ($\sim90-260$~s) the outer ribbon passes over a given pixel, a gradual rise of Mg \textsc{ii} intensity is observed (see Figure~\ref{fig:ex_lcurve}). There is a significant enhancement when the outer ribbon sits over the pixel (from comparison with the relevant SJI image). At this time the profile is enhanced, contains no central reversal, is broadened in both the red and blue wings, and the line centroid is redshifted. In some profiles there appears to be asymmetries present. Each of these features will be discussed in more detail below. 

		When the outer ribbon has moved on, the pixel sits in the inner ribbon. The intensity drops quickly from the peak, usually within one 43~s timeframe, followed by a slower decay over several minutes. The resonance lines are affected by the filament spreading over the ribbon, which results in a sudden dip in intensity in the lightcurves. The subordinate lines are seemingly not affected by the 
filament, showing instead a smooth decay to pre-flare intensity.  The profiles are still broadened at this stage though to a lesser extent. 
	
		Figure~\ref{fig:ex_flarespectra} shows the pre-flare (red line), and flaring (blue line) spectrum for a sample pixel within the NR, as well as the quiet Sun (green line) spectrum, shown previously as a reference. This figure demonstrates the spectral variations described above. It is also clear from these spectra that the subordinate lines are in absorption in the quiet Sun, and go into emission during the flare. We refer to the 2796.10\AA\ line as `s1' and since the 2798.75\AA\ and 2798.82\AA\ lines are blended we refer to them together as `s2'. The lightcurves of the h \& k lines, and of s1 \& s2, from a flaring pixel are shown in Figure~\ref{fig:ex_lcurve}, where intensity has been integrated over the line. Both the subordinate lines and resonance lines decay within a similar timescale.

	\begin{figure}
			\centering
			\includegraphics[width=.5\textwidth]{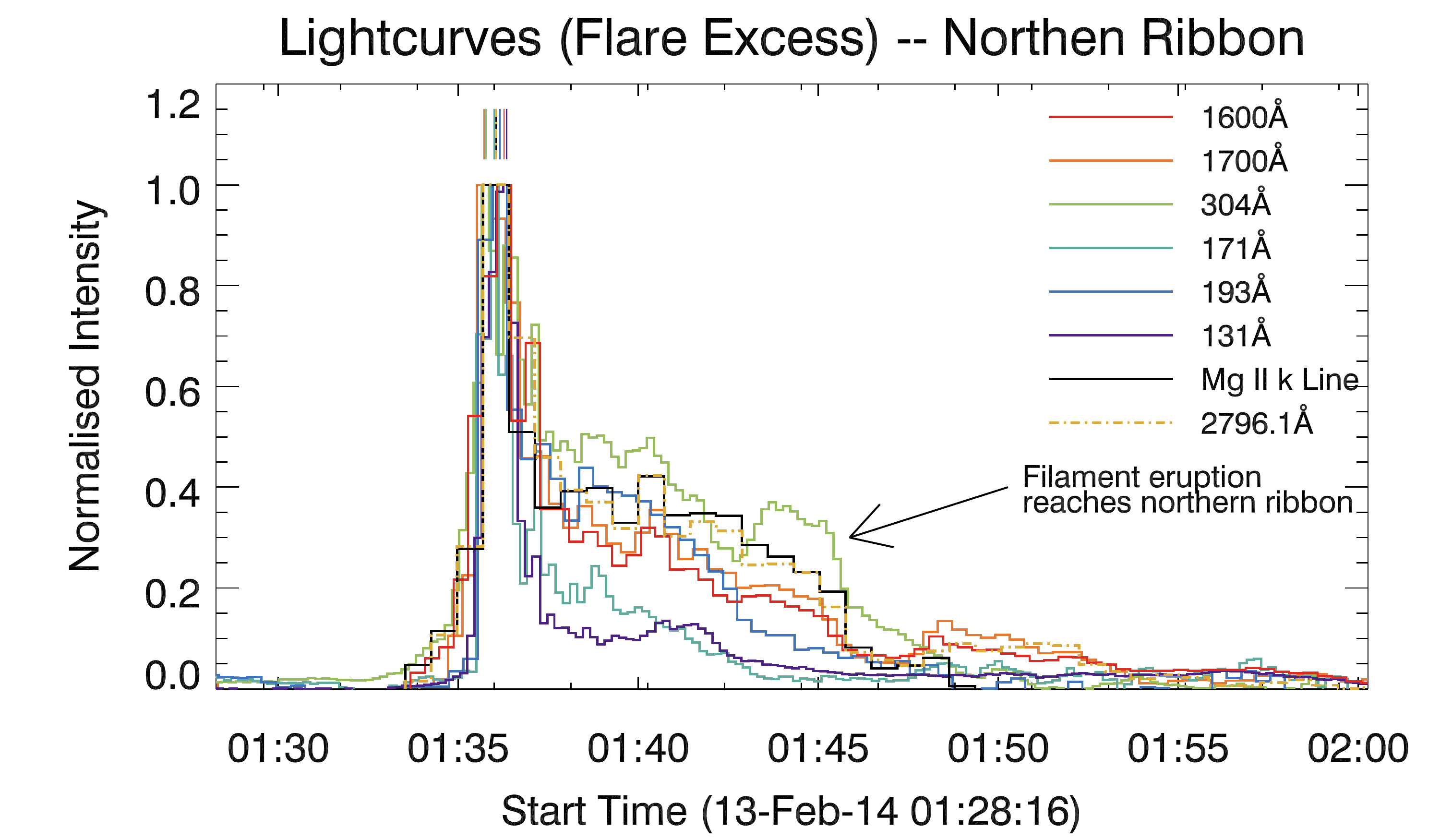}
			\caption{\textsl{Lightcurves of Mg~\textsc{ii} k and 2796.10\AA\, alongside lightcurves from various AIA channels. These are averaged over  a small source within the flaring region. The vertical lines indicate the peak times.}}
			\label{fig:ex_lcurve_aia_mg}	
		\end{figure}
		
		The k-line integrated intensity of each pixel in the NR and surrounding region is shown as a function of time in Figure~\ref{fig:int_intensity_sequence}. The pixels were stretched on their x-axis to fill the $\sim 2''$ between each slit. Since the exposure time was $\sim$ 5s each raster image takes $\sim$ 40s to produce (with times indicated on each panel). The most intense emission occurs towards the edge of the flaring region, with a particularly strong source present in the western edge at around 01:35:30~UT which is spatially associated with the RHESSI sources. Variations in the intensity occurs on sub-arcsecond scales (recall that the pixel scale in the y-direction is 0.167"). The most intense emission (the outer ribbon) is around 1-1.5$''$ wide (sometimes smaller). Additionally, some locations show repeat emission, resulting in double peaked lightcurves (separated in time by a few minutes).

		Figure~\ref{fig:int_intensity_sequence} illustrates the similar morphology of the flaring Mg~\textsc{ii} sources to observations from AIA. All passbands have a very intense source in the vicinity of the 6--9~keV RHESSI source. To investigate any temporal correlations we computed normalised lightcurves averaged over several small sources, a typical example of which is shown in Figure~\ref{fig:ex_lcurve_aia_mg}. To obtain the Mg~\textsc{ii} k-line and 2796.10\AA\ lightcurves, the integrated intensity was averaged over a flaring source $\sim$ 1.8$''$) in the y-direction along slit \# 6. AIA lightcurves were averaged over a similar size in the y-direction but slightly larger in the x-direction due to their larger pixel scale. These lightcurves show that all of the emissions are well correlated, following a similar timescale for rise, peak and decay. It is interesting to note that the chromospheric observations (1600\AA, 1700\AA, 304\AA\, k-line and 2796.10\AA) all seem to exhibit a  gradual rise time of up to several minutes before a significant enhancement to the peak intensity. By comparison the hotter lines are more impulsive.
				
				\subsection{k/h Intensity Ratio}\label{sec:khratio}

		The h \& k lines are transitions to a common lower level, from finely split upper levels. Following the arguments laid out in \cite{1997ApJ...477..509S}, \cite{1999A&A...351L..23M}, and references therein, the intensity ratios of these lines can be used to investigate their opacity.
	
		The integrated intensity of a line transition from upper level $j$ to lower level $i$ is dependent in part on the collision strength for that transition, $\Omega_{ij}$, where 
	
		\begin{equation}
			\Omega_{ij} = \frac{8\pi}{\sqrt3}\frac{I_H}{\Delta E_{ij}}gw_if_{ij}
		\end{equation}
			
		with hydrogen ionisation energy, $I_H$, threshold energy for the transition, $\Delta E_{ij}$, gaunt factor, $g$, statistical weight of level $i$, $w_i$ and oscillator strength, $f_{ij}$. Under the optically thin assumption, where the escape probability of a photon is unity, the intensity ratio of the k to h line, $R_{kh}$, is simply the ratio of the collisions strengths. For the Mg \textsc{ii} resonance lines where the transitions involve the same element in the same ionization state, to a shared lower level (so that $w_i$ is the same), and where the transition energy is similar, then the intensity ratios reduce to the ratio of the oscillator strengths, 2:1. In the optically thick case, this ratio is smaller. 
		
		\begin{figure}
			\resizebox{\hsize}{!}{\includegraphics{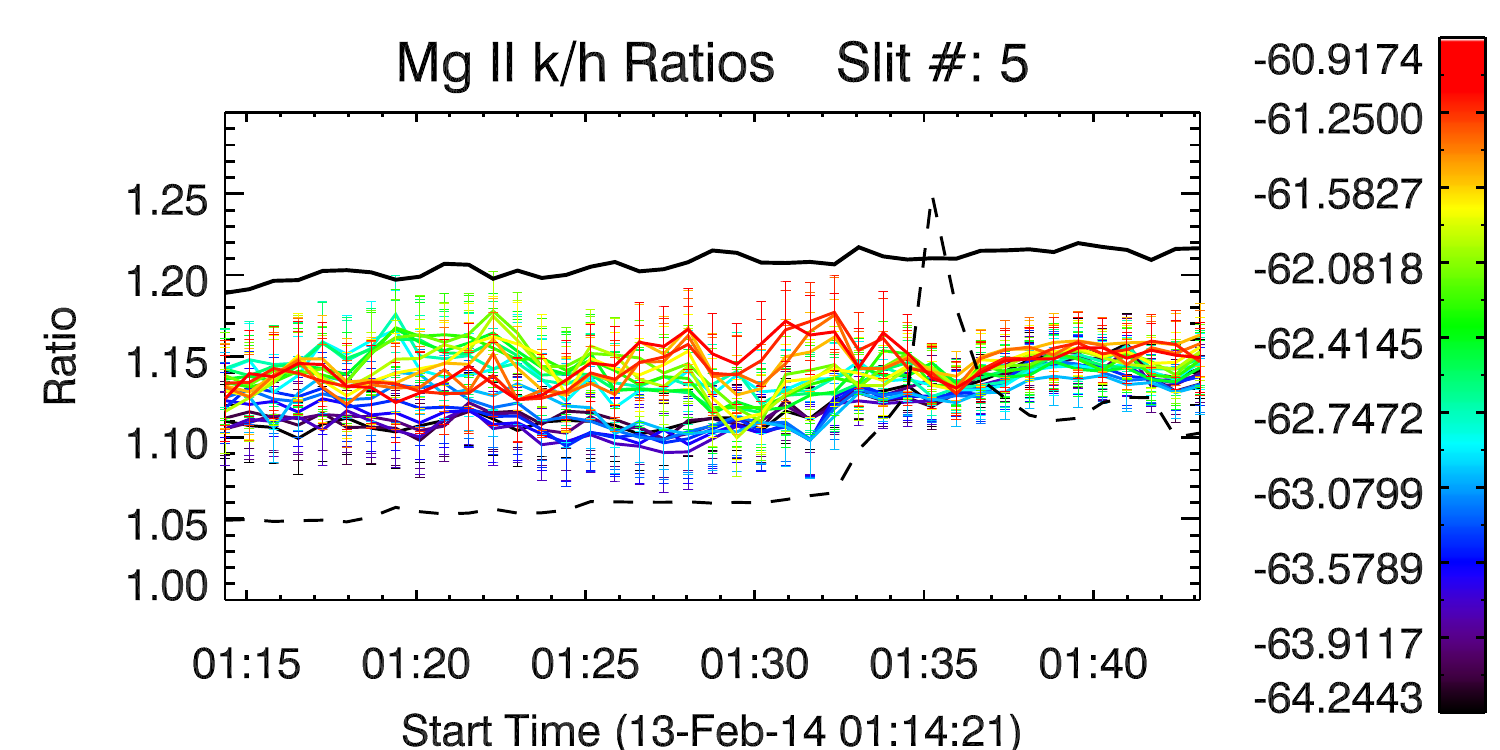}}
			\caption{\textsl{The k/h line intensity ratios for several flaring pixels (coloured lines) at slit position 5 with the quiet Sun value shown (black line at top) and a sample lightcurve to show flare times (dotted line)}}
			\label{fig:Exkhratio}
		\end{figure}

	\begin{figure*}
			\includegraphics[width=17cm]{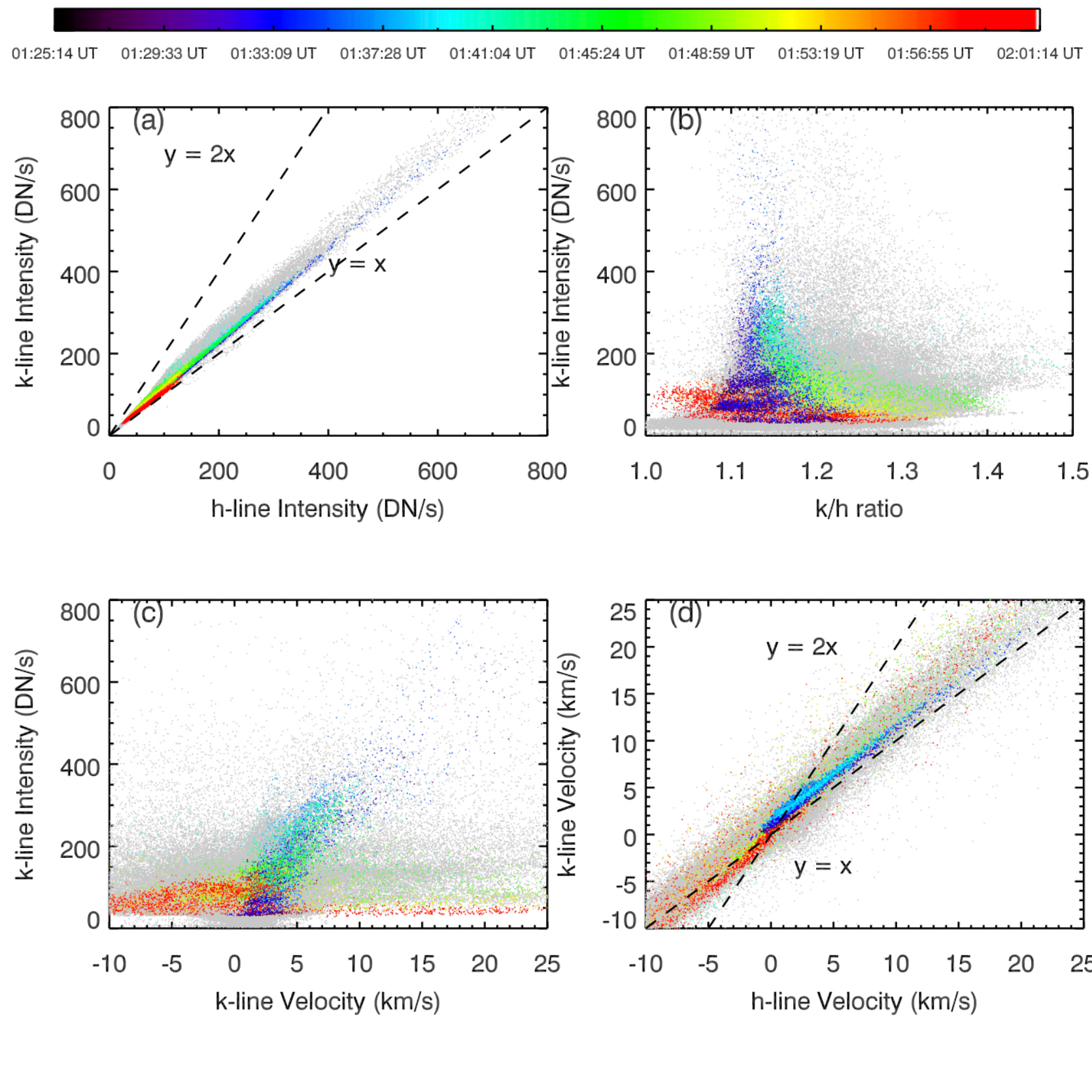}
			\caption{\textsl{Coloured pixels are pixels within the flaring region, where time refers to time. The background are pixels from the full field of view. (a) The correlation between the h \& k line intensities with lines $R_{kh} = 1$, and $R_{kh} = 2$ shown. (b) The scatter plot of $R_{kh}$ with k-line intensity. (c) The scatter plot of the k-line intensity and k-line centroid velocity. (d) The correlation between the k-line and h-line centroid velocities}}
			\label{fig:correlations}
		\end{figure*}

		For the pixels highlighted in Figure~\ref{fig:ex_sji_spectra}(b) the quiet Sun intensity ratio was measured as a function of time. These were then averaged over time, giving an intensity ratio, with standard deviation, of $R_{kh} = 1.204\pm0.010$, in line with previous studies of non-flaring sources, e.g. \citealt{1976ApJ...205..599K,1981A&A...103..160L, 1984SoPh...90...63L}) who measured a range $R_{kh} \sim 1.14-1.46$ depending on the feature observed and location on the disk.
	
		The intensity ratio was $R_{kh} \sim 1.07-1.19$ in the region surrounding the flaring emission before, during, and after the flare. Figure~\ref{fig:Exkhratio} shows $R_{kh}$ values for several pixels from slit position 5 as a function of time, with the quiet Sun value also shown, and a sample lightcurve shown in dotted lines for context.  Pixels in this region have $R_{kh}$ values less than the quiet Sun average. Comparison to AIA 1700\AA\ images show that some of the flaring pixels are co-spatial with dark pores. The only times at which these pixels show higher intensity ratios is when filament expands over the field of view (the filament material has a much higher ratio of between $R_{kh} \sim 1.4-1.7$). \cite{2014ApJ...792...93H} report an increase in the ratio following a filament eruption, for off-limb observations. 

		Although there does not appear to be a systematic increase or decrease in the $R_{kh}$ in the flaring pixels, in two slit positions there is a response to the flare. Slit positions 5 \& 6 show $R_{kh}$ converging during the flare with the spread in values decreasing (the standard deviation of $R_{kh}$ values at different cuts through each of these slit positions decreases during or shortly after the flare). Figure~\ref{fig:correlations} shows the correlation between the k \& h line intensity, with lines representing $R_{kh} = 1$ and $R_{kh} = 2$. The coloured points represent the flaring region only (defined by being between $y$ = [-60.9, -65.0]$''$) where colour refers to time. This plot shows the much tighter correlation between the k \& h line intensities in the flaring region as a whole, but particularly during the flare, in comparison to the full field of view. The effect of the filament can be seen as an increase in the ratio, and reduction in intensity, of the points at around 01:46~UT. The red/yellow points show that at later times the ratios begin to return to their previous values (the spectra begin to return to their regular profiles at these times also).

		\subsection{Quartiles Analysis of Line Profiles}\label{sec:quartiles}
		As discussed in the preceding Sections, the Mg \textsc{ii} resonance lines are optically thick and show complex features, particularly in the quiet Sun, not meaningfully fit by a Gaussian. Additionally it is not known if the line diagnostics described by \cite{2013ApJ...772...89L}, \cite{2013ApJ...772...90L} and \cite{2013ApJ...778..143P} are applicable in active region or flaring conditions. We will adopt a non-parametric approach, using quartiles to characterise the observed profiles. This is a simple, yet robust, statistical method that makes no assumption about the underlying distribution, and allows the general line properties during the flare to be quantified.

\begin{figure}
			\resizebox{\hsize}{!}{\includegraphics{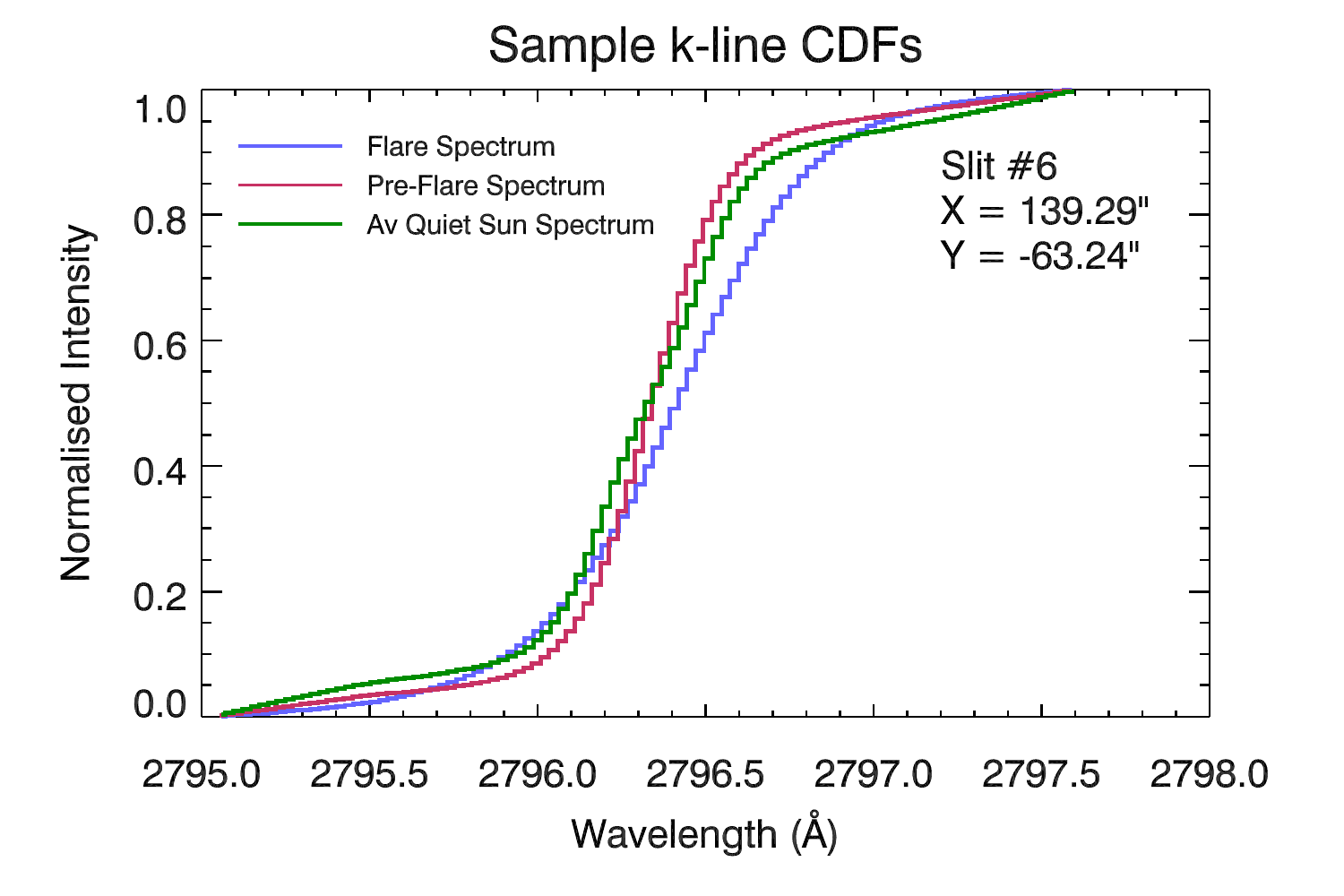}}
			\caption{\textsl{Example CDFs for a flare profile, pre-flare profile, and the average quiet Sun profile.}}
			\label{fig:ex_quarts}
		\end{figure}

		Normalised cumulative distribution functions of intensity vs wavelength (CDFs) of the h \& k lines for each profile were created (see Figure~\ref{fig:ex_quarts}) and the wavelengths corresponding to the 25\% ($Q_1$), 50\% ($Q_2$) and 75\% ($Q_3$) quartiles were found. This split each of the h \& k lines into four equal areas, with the following properties derived from the quartiles:

		\begin{enumerate}
			\item $\lambda_c = Q_2$, the line centroid wavelength \\
			\item $W = Q_3 - Q_1$, a measure of line width\\
			\item $S = \frac{(Q_3- Q_2) - (Q_2 -Q_1)}{Q_3 - Q_1}$, a measure of asymmetry. 
		\end{enumerate}
	
		The evolution of these properties with time during the flare is discussed below.
	
		\subsubsection{Centroid Motion}\label{sec:losvel}
		
		\begin{figure}
			\resizebox{\hsize}{!}{\includegraphics{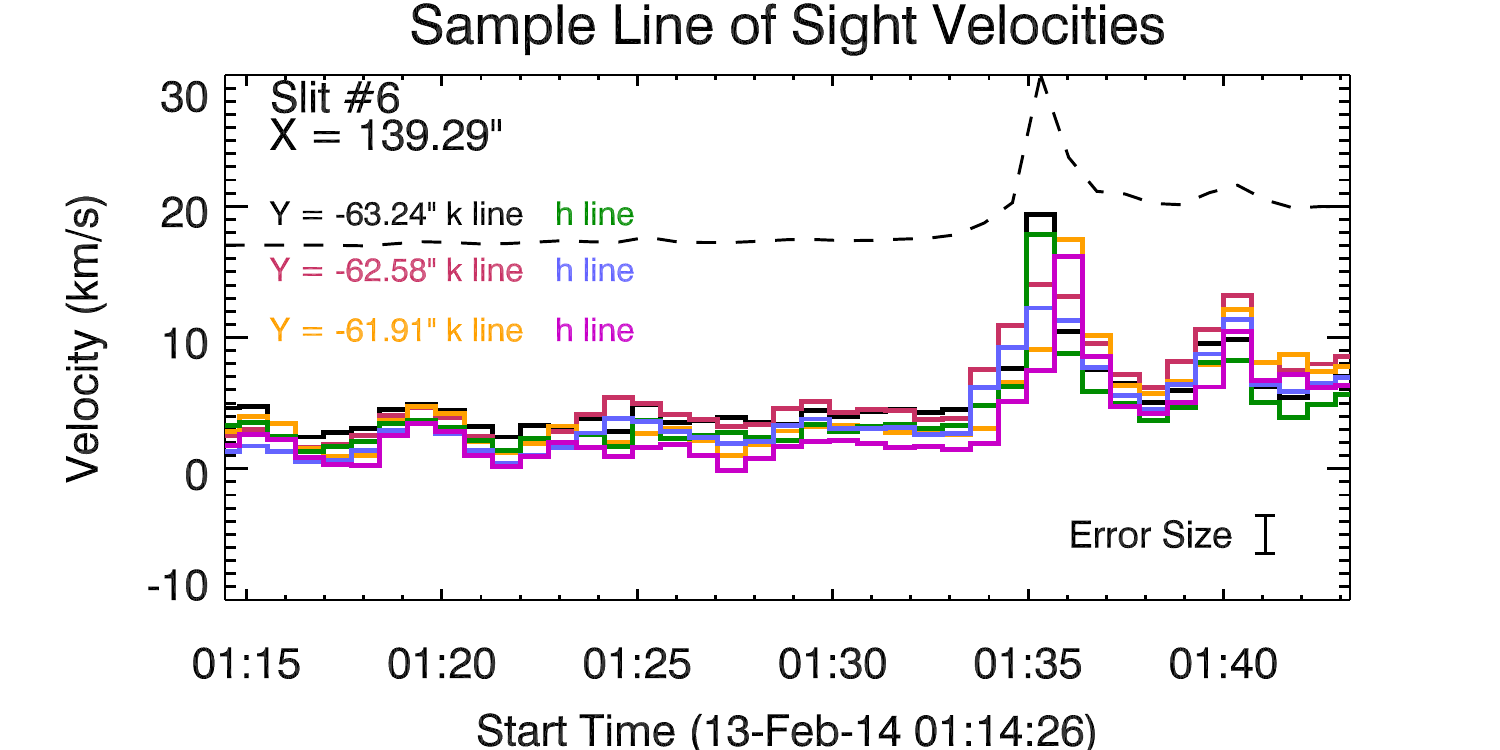}}
			\caption{\textsl{Example line of sight velocities, showing both the h and k line velocities. The dotted line is a scaled sample lightcurve, illustrating that velocity increases are associated with intensity enhancements.}}
			\label{fig:ex_vels}
		\end{figure}

		Visual inspection of the flaring spectra show that the h \& k lines have an overall redshift at the times when the outer ribbon passes over the pixel. The line of sight velocity that this redshift would be equivalent to was computed using the standard Doppler shift equation, $v_\mathrm{los} = \frac{\lambda_{c} - \lambda_{0}}{\lambda_{0}}c$, for $\lambda_{0}$ the rest wavelength, and $\lambda_{c}$ the measured line centroid. We emphasise that, since this is an optically thick line, we are measuring an equivalent velocity. It remains to be seen from modelling if this maps directly to the velocity of the atmosphere.
			
\begin{figure*}
			\includegraphics[width=17cm]{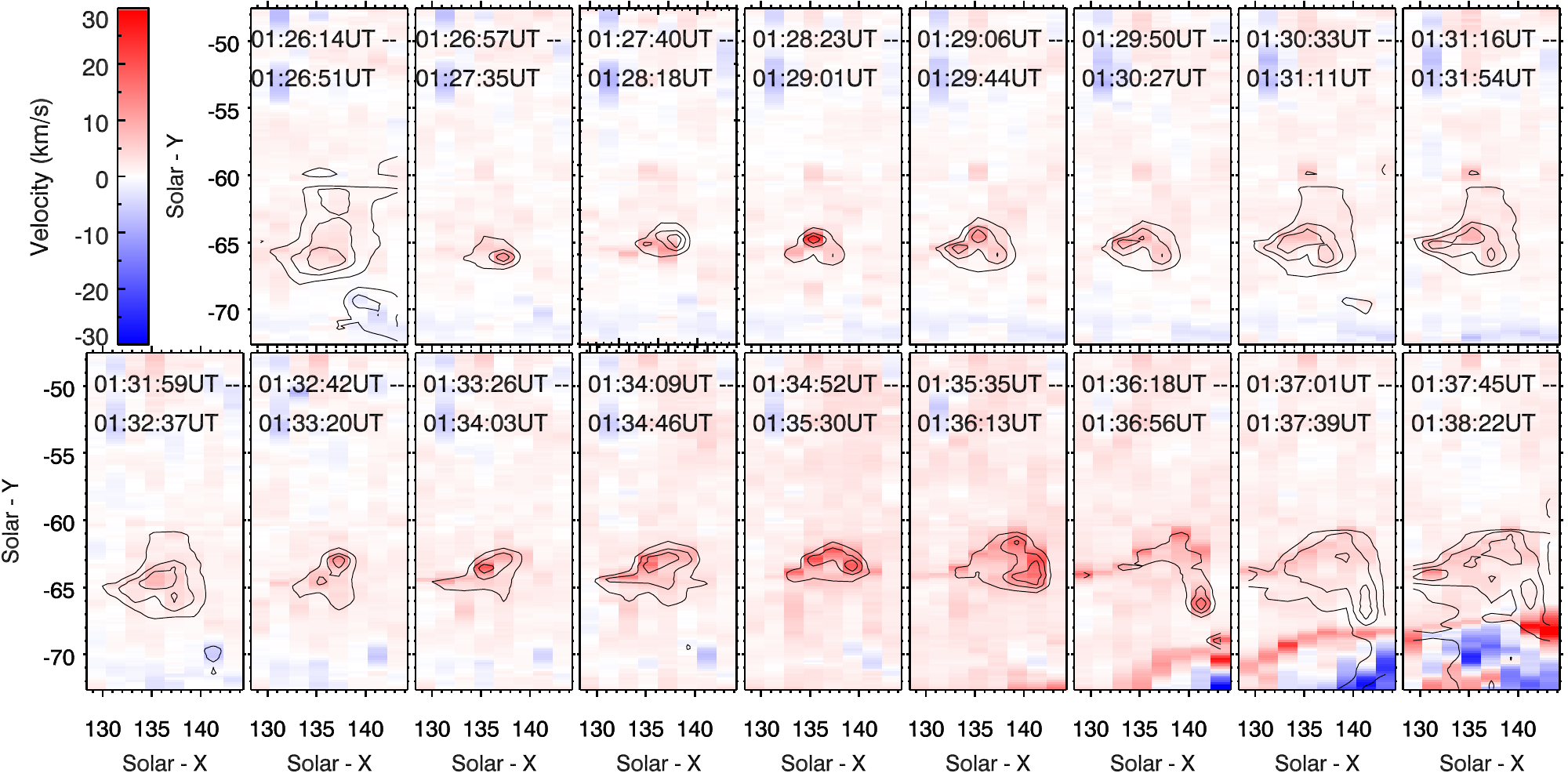}
			\caption{\textsl{The velocity of the h line over time, showing the shift of line centroid. The redshifts occur at locations of highest intensity enhancement and track the outer edge of the ribbon. The 40$\%$, 60$\%$ \& 80$\%$ intensity contours in each panel are shown for comparison.}}
			\label{fig:ex_vels_sequence}
		\end{figure*}

		Before the flare these velocities are small, and averaging over the flaring pixels in slit positions 3-7 gives a pre-flare velocity of $v_\mathrm{k,pf} = 2.28$~km~s$^{-1}$ and $v_\mathrm{h,pf} = 1.32$~km~s$^{-1}$, for the k and h lines respectively. During the flare, the line centroid is redshifted by $\sim$15-26~km~s$^{-1}$. Figure~\ref{fig:ex_vels} shows the h \& k velocities from some sample pixels from slit position 6, and Figure~\ref{fig:ex_vels_sequence} shows a map of the h-line centroid velocity for the northern ribbon. The coloured lines in Figure~\ref{fig:ex_vels} show the k and h line velocities of several pixels as a function of time, which show a clear response to energy input during the flare.  Contributions to the error on $v_\mathrm{los}$ were from the error on rest wavelength and on the centroid wavelength. The dominant source of error on the centroid wavelength measurement was due to the absolute wavelength calibration of the NUV channel, which \cite{2014SoPh..289.2733D} state as being $\pm$1~kms$^{-1}$. Errors on the rest wavelength were a combination of the absolute wavelength calibration, and the standard deviation of the rest wavelength measurements from averaging over quiet Sun pixels ($\pm$ 0.0032\AA\ = 0.343~kms$^{-1}$). The size of the error on $v_\mathrm{los}$ is shown in the lower right corner of this figure. The dotted line shows a sample k-line integrated intensity lightcurve, with intensity peaks matched by peaks in velocity. However, the velocity enhancement decays quicker than the intensity, decaying by half within one timeframe (43~s), and then almost to the background within 3 minutes, whereas the intensity takes several minutes to decay to background levels. If we use the subordinate lines as a proxy for the resonance lines (as mentioned above, they share a similar temporal profile, but the subordinate lines are unaffected by the filament), then the time to decay to pre-flare levels is over 15 minutes, considerably longer than the redshift lifetime. This can be also be seen by comparing the ribbon structures in the intensity maps (Figure~\ref{fig:int_intensity_sequence}) with the velocity maps shown in Figure~\ref{fig:ex_vels_sequence}. The velocity maps show the outline of the outer ribbon (the most intense emission in the intensity maps), whereas the inner ribbon returns to close to the background much quicker. The velocity curves also mimic the gradual rise in intensity before the main peak. This suggests that the centroid redshift is associated with the deposition of energy into the pixel, and that the magnitude of the velocity and intensity are related. 
		
		Figure~\ref{fig:correlations}(c) shows the correlation of the k line intensity and the k line centroid velocity. Here dark blue to light green represents the flaring times, and a trend for higher redshifts with higher integrated intensity is apparent. There is a spread in this feature of the plot, though, suggesting that while a higher intensity is associated with a higher redshift, there are other effects present. The green and red points are the times at which the filament expands over the northern ribbon, producing the large blueshift at later times.
		
		We note that while the h \& k lines follow the same trends and are enhanced in the same locations, the k line velocities are consistently higher than the h line velocities (albeit by a small amount). This is true of both the quiet Sun, pre-flare and flare profiles. Figure~\ref{fig:correlations}(d) shows the correlation between the h line and k line centroid velocities. The magnitude of the k line velocity is mostly greater than the h line velocity. It must be noted, however, that while the k-line centroid velocity is systematically larger than the h-line, the magnitude of this difference lies within the error bar size.

		\subsubsection{Line Widths}\label{sec:linewidth}
		The measure $W = Q_{3} - Q_{1}$ describes the width of the spectral line that is bounded by the 25th and 75th percentiles, analogous to the FWHM. 	
		As the flare begins, the width of the line increases to approximately $W \approx 0.45-55$\AA\, with a maximum width during the flare of $W = 0.623$\AA\ in the k-line, and $W = 0.561$\AA\ in the h-line. As in the previous case, the line core intensity has increased, but here we are also observing enhancement to the line wings. This broadening, which is temporally correlated with the lightcurves, suggests that the lower layers of the chromosphere are also affected. The width of the flaring pixels remains enhanced immediately following the flare but because the filament material expands over the region, it is not possible to determine whether this ongoing broadening is due to the flare.

		\subsubsection{Asymmetry}\label{sec:lineskew}
		Asymmetries in the line profiles could provide additional information about plasma motions. Using the measure $S = \frac{(Q_3-Q_2) - (Q_2-Q_1)}{Q_3 - Q_1}$ the asymmetry around the line core was investigated. This measure looks at the positions of the 25\% and 75\% quartiles relative to the line centroid, so that if there is a greater enhancement redward of the line core versus blueward $S$ will return a positive number, and a negative number for a greater blueward enhancement.
		
		The region immediately surrounding the flaring material was observed to have a small red asymmetry that occasionally became a blue asymmetry, whereas the samples from a wider region showed much stronger red asymmetry. Both cases had a lot of variation over time. 
	
		During the flare, however, the most intense emission (around 01:35 -- 01:37~UT) had a clear blue asymmetry associated. As can be seen in Figure~\ref{fig:skew_sequence}, the asymmetry shortly before this intense emission was around zero in the ribbon, and positive in the area surrounding the ribbon, but in the panel representing times 01:34:52--01:35:30~UT the ribbon outline becomes apparent with a blue asymmetry. This blue asymmetry strengthens over the next few minutes, and at all times is spatially correlated with high intensity emission. By 01:37~UT this asymmetry has reduced greatly to the pre-flare values. The flaring spectrum shown in Figure~\ref{fig:ex_flarespectra} shows an evident blue asymmetry. 
		
		\begin{figure}
			\resizebox{\hsize}{!}{\includegraphics{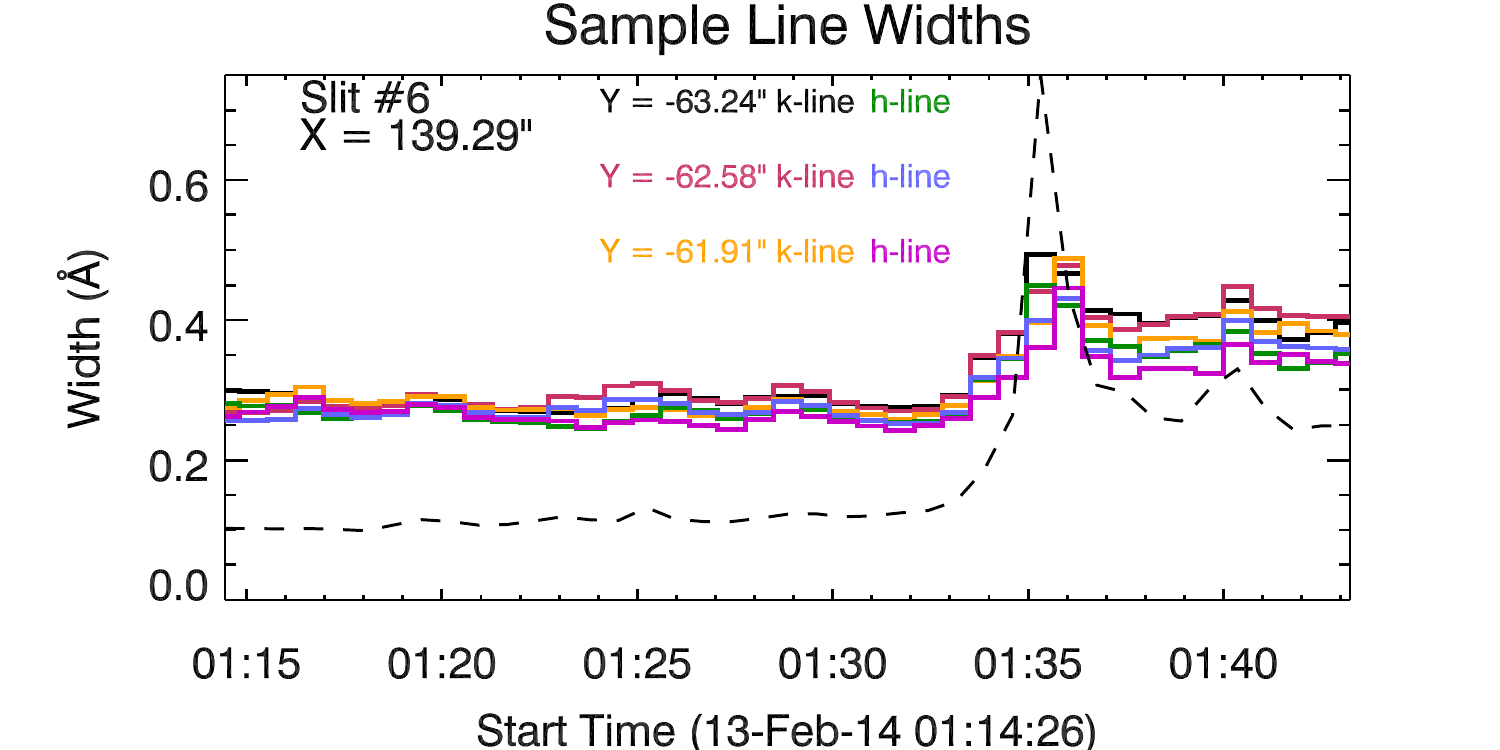}}
			\caption{\textsl{Width as a function of time for several pixels in the flaring region. The dotted line is a scaled lightcurve illustrating the response of the line width to intensity enhancements.}}
			\label{fig:ex_widths}
		\end{figure}


	\section{Discussion}\label{sec:discuss}
	The flaring Mg \textsc{ii} resonance line profiles are very different from quiet Sun spectra, and show a clear response to the changing atmosphere following the flare. The central reversal feature (k3) that is almost ubiquitous in the quiet Sun is absent in the flaring profiles, yet the k:h line intensity ratio suggests the lines are still optically thick. This behaviour has been seen in sunspot umbras \citep{1982ApJS...49..293L, 2001ApJ...557..854M} though umbral models have so far failed to produce synthetic spectra that match observations (umbral models have only produced self reversed line cores, unless an optically thin assumption is used). However, the flare ribbon that we studied is not located entirely over a sunspot umbra. It is situated close to a pore, but most of the ribbon structure does not sit over the pore.  

	Optically thick resonance line formation is a very complex physical process that requires advanced radiative transfer modelling to fully interpret emergent profiles, but we are able to state that during the flare the Mg \textsc{ii} h \& k line source functions must increase with height to produce emission at line centre. Usually, in the case where profiles are reversed, the source function increases and is still partially coupled to the Planck function until it reaches a maximum, producing the k2/h2 peaks. Following this point, it decreases in the upper atmosphere, producing the k3 minimum.
	
	Previous observations of flares using the Ca \textsc{ii} H \& K lines (with a similar formation process) have shown that the line centre can go into emission. \cite{1968AJS....73Q..69M}, following theoretical work on resonance line formation by \cite{1960ApJ...131..695J} and \cite{1968SoPh....3..181A} used an assumption of a source function with frequency independence and complete redistribution (CRD, which is approximately valid over the core of the resonance lines, within $\sim$ 3 Doppler widths of line centre) to show that an increase in density or electron temperature could account for these observations. They wrote this frequency independent part of the source function as

	\begin{equation}
		S = \frac{\int J_{\nu}\Phi_{\nu}d\nu~+~ \epsilon B_\nu(T_e)}{1~+~\epsilon}
	\end{equation}

	where $J_\nu$ is the mean intensity, $\Phi_\nu$ is the normalised absorption profile, $B$ is the Planck function and $\epsilon = C/A$ is the ratio of collisional to radiative de-excitation. The coupling of the source function to the Planck function contributes to the line profile of the Ca \textsc{ii} (and Mg \textsc{ii}) resonance lines. The source function departs from the Planck function and reaches a maximum in the middle chromosphere where the emission peaks are formed. It then decreases with increasing height reaching a minimum where the line centre is formed (c.f \citealt{2013ApJ...772...89L} Figure 7). If $\epsilon$ increases in the mid-upper chromosphere due a flare-induced higher density ($\epsilon \propto$ n$_{e}$) then this combined with a temperature increase could lead to a source function that increases with height, producing a core in emission. This discussion involved a simplified model of the radiative transfer,  but the main arguments should hold. An increase in the density will couple the line source function to the Planck function more strongly, which would result in a more intense line profile with a core in emission if the source function continues to increase with height at the $\tau = 1$ level of the line core. 
	
		\begin{figure*}
			\includegraphics[width=17cm]{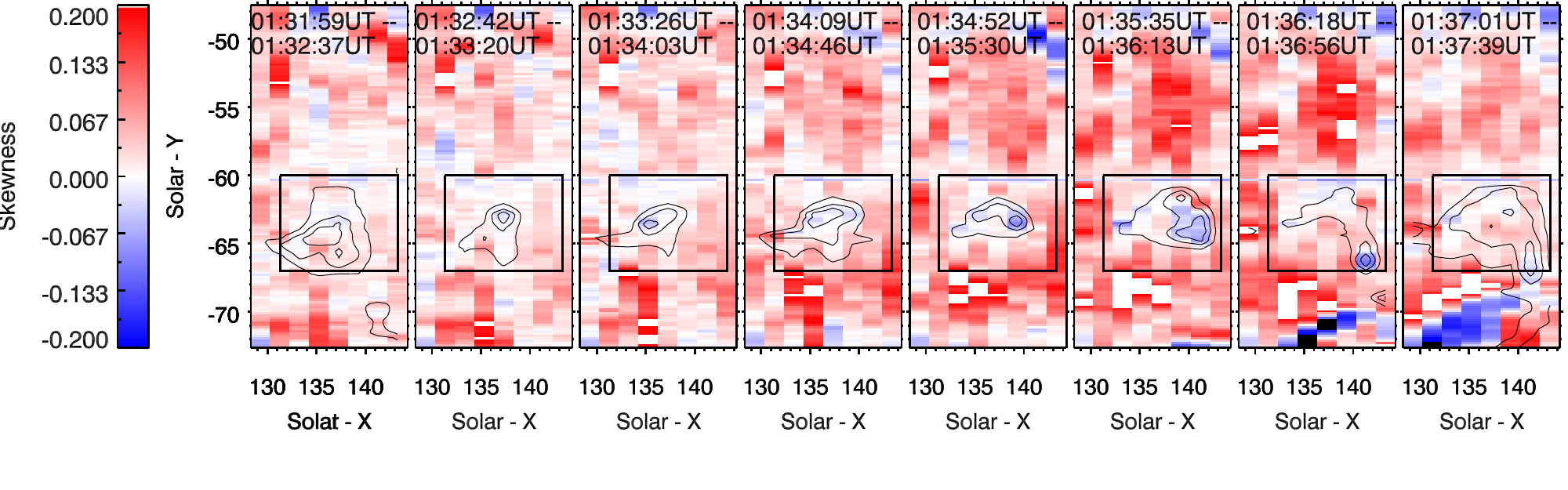}
			\caption{\textsl{The asymmetry in the k-line (h-line similar). The box shows the location of the northern ribbon, where a blue asymmetry can be seen in several pixels between $\sim$01:34~UT -- $\sim$01:37~UT. The 40\%, 60\% \& 80\% intensity contours in each panel are shown for compairson.}}
			\label{fig:skew_sequence}
		\end{figure*}

	Electron impact excitation is the dominant mechanism to populate the resonance and subordinate line upper levels in the quiet chromosphere, a process which is sensitive to density and temperature. We can therefore state that a higher intensity line is indicative of a density enhancement. Other mechanisms for populating the resonance line upper levels might complicate matters further, removing this sensitivity to density. A high lying level of Mg~\textsc{ii} can be excited by Ly$\beta$ pumping, leading to cascades down to the h \& k line upper levels. This process was found to be negligible in the quiet chromosphere by \cite{1983A&A...125..241L} and preliminary results suggest it is also negligible in flares (J. Leenaarts 2015, \textsl{private communication}). 

	In addition to the resonance lines, we observed a significant enhancement to the 3p-3d level subordinate lines. These lines normally appear as absorption features, but went into emission during the flare. Recent modelling work using the Bifrost code of \cite{2011A&A...531A.154G} and the radiative transfer code, RH \citep{2015A&A...574A...3P} investigated the formation of these lines \citep{2015ApJ...806...14P}. They found that these optically thick lines are formed in the lower chromosphere, and that they only go into emission when there is a large temperature gradient ($\geq$ 1500~K) between the temperature minimum and formation region (a column mass of 5$\times$10$^{-4}$~g/cm$^{3}$; the lines were not sensitive to an increased gradient above this column mass). 

	In the context of our flare observations, significantly enhanced subordinate line emission is an indicator that a large temperature gradient could be present, and that flare related heating has reached very low layers. The integrated intensity of the subordinate lines behaves like that of the resonance lines. The subordinate line intensity began to rise slowly over a period of 2-4 minutes followed by a sharp increase to a peak. They take a similar length of time to return to background levels as the resonance lines. Subordinate lines do not appear in emission in filament spectra, only in flaring spectra. In a future study, we plan to investigate the intensity, width and centroid shift of the 2796.10\AA\ subordinate line (the other lines are blended, complicating their study), in relation to the resonance lines, as this would sample these properties in different atmospheric layers.  \\
		 
	The ratio of k-line to h-line intensity ($R_{kh}$) suggests optically thick emission, throughout all stages of the flare, and is a similar value to those measured by \cite{1984SoPh...90...63L}. As Figure~\ref{fig:Exkhratio} shows, there is a variation in the value of $R_{kh}$, with values ranging from $\sim$1.07-1.19 when considering the errors on individual points. Variations are present in both time and spatial location, and a similar pre-flare variation is seen across all slit positions. It is worth noting that \cite{1984SoPh...90...63L} reported much larger variations in this value (between 0.9 and 1.5) though with much less spatial resolution that IRIS provides. Around the time of the flare the spread of values drops quite clearly in two slit positions (5 \& 6). We speculate that this could be due to differences in heating during the pre-flare atmosphere and the flaring atmosphere. Small scale variations in heating or energy transport in the non-flaring chromosphere could cause a spread in the value of $R_{kh}$. These slit positions are adjacent, and are located where the majority of the flare energy is deposited. We also note that these slit positions lie close to the compact sources observed by RHESSI.

	It is clear from the profiles, and our analysis of the line of sight velocities, that the flare causes an overall redshift of both the Mg \textsc{ii} resonance lines. \cite{2013ApJ...772...90L} found that the line of sight velocity of the k$_3$ and h$_3$ components match very well to the physical velocity of the chromosphere at their formation height. Since we are dealing with very different atmospheric conditions, we are not able to state that the observed line of sight velocities map directly to the velocity of the upper chromosphere. Modelling of this line in a flare atmosphere should indicate if a direct mapping of centroid velocity to upper atmospheric velocity exists. A downward motion is expected in models of explosive chromospheric evaporation \citep{1985ApJ...289..434F, 2005ApJ...630..573A}, and has been observed previously in solar flares, including the optically thick He \textsc{ii} line which showed a similar magnitude of centroid shift as we observe in Mg \textsc{ii} (see \citealt{2009ApJ...699..968M} and references therein). \cite{1980IAUS...91..217S} and \cite{1995SoPh..158...81D} discuss finding similarly shortlived downflows that were only associated with the outer edge of the ribbon.
	
	 As well as the overall redshift of the line, we find a strong blue asymmetry in certain locations within the ribbon. These locations are the sources of strongest emission. Similar observations in other wavelengths have been reported before (\citealt{1994SoPh..152..393H}), but usually  associated with the initial stages of the flare. An interpretation for asymmetry in the spectral lines is that there are both upflows and downflows present in within one pixel, with a blue shifted contribution to the line superimposed onto the more intense redshifted contribution. An alternative explanation, as suggested by \cite{1994SoPh..152..393H} is that this blue asymmetry is a signature of \textsl{downward} propagating plasma which absorbs radiation from the red side of the line core, making the blue wing appear more intense and causing a short lived blue asymmetry. This has to be investigated by modelling.
	 
	During the flare the resonance lines become significantly broadened over both the quiet Sun and pre-flare profiles. During the flare both the line width, $W$, and the base width increase significantly, with a similar temporal profile to the lightcurves (though they do not decrease in width as quickly as intensity does). This increasing width shows that during the flare the intensity enhancements are not confined to the line core region (i.e. k2/k3 components) but that the near wings and k1 minima are affected. 
The Doppler width ($v_{th} = \sqrt{2k_bT/m_{mg}}$) of the line is around 2--4.2~km~s$^{-1}$ (0.02--0.04\AA) for typical chromospheric temperatures. Even if we assume that flare footpoints in the chromosphere are significantly heated, it is unlikely that the increased width can be accounted for by a temperature increase. As we have demonstrated, the Mg \textsc{ii} lines are optically thick, even during the flare, and so opacity broadening is important, with photons scattered into the inner wings. Flare effects may results in additional scatterings to the wings, broadening them over the quiet Sun/pre-flare widths. Turbulence and unresolved flows may act to widen the lines. Although the line centroids have an overall redshift, plasma moving at different velocities would broaden the profile.

	
	\section{Conclusions}
	We have analysed Mg \textsc{ii} h \& k line observations during a solar flare. The Mg~\textsc{ii} intensity enhancements were well correlated with enhancements in other UV passbands as observed by AIA. We also reported the RHESSI HXR observations around the peak times, to guide any subsequence modelling of this event. RHESSI observations showed a largely thermal event, with compact thermal sources that were spatially associated with the flare ribbon. Hard x-ray counts above 20~keV were only present towards the peak of the event (though remained weak), and non-thermal sources were only spatially associated with flare ribbons in one image. 
	
		During the flare the central reversal in the Mg~\textsc{ii} h \& k lines vanishes but analysis suggests the lines are optically thick. The spatial and temporal evolution of Mg \textsc{ii} during the flare shows very intense, spatially localised energy input at the outer edge of the ribbon (with variations on scales of $\sim$0.5$''$ or less). There is evidence of a slow onset of excitation before a main impulsive peak. This energy deposition results in pixels within the ribbon that show the following properties:

	\begin{enumerate}
		\item Line centroids that are redshifted with an equivalent velocity of $\sim$15-26~km~s$^{-1}$ 
		\item Broadened line profiles, with the width increasing from $W\sim0.28\AA\ $to $W \sim 0.45-55$\AA\ 
		\item Blue asymmetries in intense pixels.
		\item The k/h intensity ratio becoming more uniform in two slit positions, with $R_{kh}\sim1.15$
	\end{enumerate} 

	These observations should be compared to synthetic flare spectra in order to test flare models and improve our understanding of the atmospheric response to flare energy input. It is our intention to use RADYN, the radiation hydrodynamic code of \cite{2005ApJ...630..573A}, to model the atmospheric response to flares, and to input these atmospheres to RH, the advanced radiative transfer code of \cite{2015A&A...574A...3P}, to compute the flaring Mg \textsc{ii} spectra.

	\begin{acknowledgements} 
	The authors would like to thank Drs S. Jaeggli, Y. Li and H. Tian for help concerning the use of IRIS data. Additionally we would like to thank Dr P. Judge for helpful discussions of radiative transfer, and the referee, who's comments helped to improve this work. GSK acknowledges the hospitality of the Solar Physics group at Montana State University for hosting a research visit, and the financial support for this visit via a mobility scholarship (as well as PhD funding via a postgraduate research scholarship) from the College of Science and Engineering, University of Glasgow. The research leading these results has received funding from the European Community's Seventh Framework Programme (FP7/2007-2013) under grant agreement no. 606862 (F-CHROMA). JQ would like to acknowledge funding from NASA grant NNX14AC06G.
	\end{acknowledgements}

	\bibliographystyle{aa}
		\bibliography{Kerr_etal_MgII_arXiv}

\end{document}